\pdfoutput=1
\documentclass[aps,pra,reprint,superscriptaddress,amsmath,amssymb,longbibliography,floatfix]{revtex4-2}
\usepackage[T1]{fontenc}
\usepackage{lmodern}
\input{glyphtounicode}\pdfgentounicode=1
\usepackage[utf8]{inputenc}
\usepackage{graphicx}
\usepackage{bm}
\usepackage{booktabs}
\usepackage{tabularx}
\usepackage{amsmath}
\usepackage{amssymb}
\usepackage[expansion=false]{microtype} 
\usepackage[colorlinks=true,linkcolor=blue,citecolor=blue,urlcolor=blue]{hyperref}

\newcommand{\ket}[1]{\lvert #1\rangle}
\newcommand{\bra}[1]{\langle #1\rvert}
\newcommand{\Dstate}[2]{\ket{D_{#1}^{(#2)}}}
\newcommand{\gamin}{\gamma_{\mathrm{in}}}
\newcommand{\gamsh}{\gamma_{\mathrm{sh}}}
\newcommand{\Fin}{F_{\mathrm{raw}}}
\newcommand{\Fabs}{F_{\mathrm{abs}}}

\newcommand{\Fop}{\widehat{F}_{\mathrm{RZ}}}
\newcommand{\Cnk}{\binom{N}{K}}
\newcommand{\Ucoll}{U_{\mathrm{coll}}}
\newcommand{\SWAP}{\mathrm{SWAP}}

\begin{document}

\title{Optimizer or physics? Machine-learning surrogates for collision-based preparation of multipartite entangled quantum states}

\author{Duc-Kha Vu}
\affiliation{Department of Electrical and Computer Engineering, Saint Louis University, St.~Louis, Missouri 63103, USA}

\author{Minh Tam Nguyen}
\affiliation{KT One, Tampa, Florida 33613, USA}

\author{\"Ozg\"ur E. M\"ustecapl{\i}o\u{g}lu}
\affiliation{Department of Physics, Ko\c{c} University, Istanbul, T\"urkiye}
\affiliation{T\"UB\.ITAK Research Institute for Fundamental Sciences (TBAE), Gebze, T\"urkiye}

\author{Fatih Ozaydin}
\email{mansursah@gmail.com}
\affiliation{Institute for International Strategy and Emerging Technologies, Tokyo International University, Tokyo 170-0013, Japan}
\affiliation{Data Science \& AI Major, Tokyo International University, Tokyo 170-0013, Japan}
\affiliation{Nanoelectronics Research Center, Istanbul, T\"urkiye}

\date{\today}

\begin{abstract}
Repeated partial-SWAP collisions can prepare symmetric Dicke states, including the \(W\) state, without measurement or postselection. The collision sequence uses only two global collision angles, and its fidelity is evaluated after a final layer of local phase rotations is optimized numerically, so each candidate pair requires a many-qubit simulation and a phase fit. We study when learned surrogates reduce the number of such evaluations, and what limits the best-found fidelity, for 65 Dicke targets with 11--20 qubits in noiseless simulations. At 50 evaluations per target, Gaussian-process Bayesian optimization (GP-BO) exceeds fixed Sobol and grid designs by 0.088 and 0.063 in mean best-found fidelity. Bounded Nelder--Mead reaches 0.760 against 0.762 for GP-BO, and has the higher three-seed mean on 41 of the 61 targets with a reference value; in this two-parameter problem, adaptivity matters far more than the choice between the two methods. A neural-network ensemble pretrained on 750 observations from smaller systems needs 32\% fewer new evaluations per target than GP-BO at comparable fidelity, but the full neural search takes 1.6 times the wall time of GP-BO in our implementation. In a separate source-only test, the best of five neural proposals per target exceeds the best of five population-model proposals by 0.070 in mean fidelity. Exact symmetries and state diagnostics help explain why the leading adaptive methods reach similar family-mean fidelities. The collisions conserve the magnitude of the bare Dicke overlap and the total-spin weights, and for the best controls on a 12-target diagnostic panel, 79\% of the infidelity is an amplitude deficit that a final phase layer cannot remove, which motivates interleaved phase control. 
\end{abstract}

\maketitle

\section{Introduction}\label{sec:intro}

Dicke states are permutation-symmetric states in which a fixed number of excitations is shared coherently by all qubits of a register~\cite{dicke1954}. The single-excitation member of the family is the \(W\) state. Its pairwise entanglement survives the loss of particles~\cite{dur2001,neven2018}, and it shows stronger nonclassicality than the Greenberger--Horne--Zeilinger state~\cite{sen2003}. Dicke states with more excitations have larger entanglement depth~\cite{chen2016} and remain useful for metrology in the presence of noise~\cite{saleem2024}. Preparing them reliably is a standard problem in quantum control.

Even the \(W\) state is not straightforward to prepare, since it cannot be enlarged by local operations alone. In photonics, linear-optical gates were proposed and demonstrated that expand a \(W\) state by acting on a single photon or convert entangled photon pairs into a \(W\) state~\cite{tashima2008,tashima2009njp,tashima2009prl,tashima2010}. Fusion gates join smaller \(W\) states into larger ones~\cite{bugu2013,yesilyurt2013,ozaydin2014,zang2015,li2016}, and deterministic schemes grow a \(W\) state by adding separable ancillary qubits~\cite{yesilyurt2016,zang2016,ozaydin2021}. Other routes use Pauli spin blockade in quantum dots~\cite{bugu2020} or linear optics with a quantum eraser~\cite{kim2020}. For more than one excitation, the proposals include gate-based circuits with explicit gate counts~\cite{bartschi2019,chakraborty2014,mukherjee2020,aktar2022}, mid-circuit measurement with feedback~\cite{yu2024}, global control of Ising-coupled qubits~\cite{stojanovic2023}, phase estimation in a spin ensemble~\cite{wangterhal2021}, adiabatic and counterdiabatic driving~\cite{linington2008,opatrny2016,carrasco2024,wang2026dicke}, engineered dissipation~\cite{zhu2025}, gates that transform one Dicke state into another~\cite{kobayashi2014}, and expansion when only part of the register is accessible~\cite{thapa2025}. Each approach assumes a different set of resources, such as individual addressing, measurement, ancilla qubits, or a particular native interaction.

Collision models describe a system that interacts with a sequence of ancillary units, one at a time~\cite{ciccarello2022,campbell2021}. They were developed for open-system dynamics and can simulate any multipartite Markovian evolution~\cite{cattaneo2021}, but they can also be used to build entanglement. \c{C}akmak \textit{et al.} showed that a mobile shuttle qubit can generate multipartite entanglement through excitation-preserving partial-SWAP collisions with registers that do not interact directly, and that a projective measurement on the shuttle improves the preparation of a four-qubit \(W\) state~\cite{cakmak2019}. In recent work~\cite{vu2026pra}, we eliminated the shuttle measurement. The initially excited shuttles remain part of the final state, and partial-SWAP collisions prepare multi-excitation Dicke states deterministically in principle. The protocol has two control parameters, one for collisions within the registers and one for collisions between a shuttle and a register qubit. They were optimized with multistart L-BFGS-B~\cite{byrd1995} for targets of up to 14 qubits, and the protocol was tested against missed collisions and decoherence. Reference~\cite{vu2026data} translated the resulting depth--fidelity frontiers into circuit costs and ran selected circuits on a superconducting processor. In a recent work~\cite{nguyen2026mrz}, layers of local \(R_Z\) rotations are added between collision rounds, and the reachability, quantum-geometric conditioning, and preparation cost of this larger control set are analyzed. However, none of these works used learning to choose the controls.

In this paper we treat the choice of the two collision angles as a black-box optimization with an expensive objective. The search space is two dimensional, but each candidate has to be scored by propagating an \(N\)-qubit state through up to 200 ordered collision subrounds and then fitting a final layer of local \(Z\) rotations. The difficulty lies in the number of evaluations rather than in the number of parameters, the regime for which surrogate models and Bayesian optimization were developed~\cite{shahriari2016,snoek2012}. In quantum control, Bayesian optimization has been applied when measurement statistics are limited~\cite{sauvage2020}, to higher-dimensional problems with several levels of simulation accuracy~\cite{lazin2023}, and to the online tuning of a cold-atom experiment~\cite{wigley2016}. Reinforcement learning~\cite{bukov2018} and physics-informed neural networks~\cite{norambuena2024} have been used for state preparation and control, and surrogate-guided search has lowered the simulation cost of quantum-network design~\cite{prielinger2025} and of variational circuit learning~\cite{oleary2025}. Comparisons with conventional optimizers show that the gain from learning depends on the control constraints and on the size of the problem~\cite{zhang2019rl}, and benchmarks of derivative-free solvers find that their ranking changes with dimension and evaluation budget~\cite{rios2013}. In a two-parameter problem, a surrogate has to compete with strong classical methods at the same budget, and any data it reuses from other system sizes must be counted as part of its cost.

Reference~\cite{nguyen2026mrz} shows that interleaved phase layers can bring the Dicke fidelity close to one. The passive protocol remains of interest because it needs far fewer control parameters and no single-qubit control during the collision sequence. It uses two global collision angles and a single final layer of local phases. Inserting phases after every collision round adds \(N-1\) independent parameters per round, or \(T_{\mathrm r}(N-1)\) for \(T_{\mathrm r}\) rounds, and requires individual phase control of every qubit in the middle of the sequence. On superconducting processors, local \(Z\) rotations are usually implemented as virtual frame updates~\cite{mckay2017}, so the interleaved phases add little or no gate time there, provided the native exchange gate can follow the frame changes. On platforms where single-qubit control during the sequence is slow, noisy, or unavailable, and in settings where two global parameters are much easier to calibrate or learn than hundreds, the passive protocol is the relevant minimal-control limit, and it is the subject of this paper.

We compare neural-network surrogates that pool data from several system sizes, Gaussian-process Bayesian optimization (GP-BO) run separately for each target, and classical derivative-free searches on 65 Dicke targets with 11--20 qubits, simulated without noise. GP-BO and the classical searches use 50 evaluations per target, some of them reused initial observations; the neural search allocates new evaluations adaptively across targets and also draws on 750 observations from smaller systems. We account separately for new target evaluations, reused data, and wall time. The main findings are the following.
(i)~GP-BO is ahead of fixed Sobol and grid designs by 0.06--0.09 in mean fidelity after 50 evaluations.
(ii)~Bounded Nelder--Mead ends within 0.002 of GP-BO and has the higher target-level mean on 41 of 61 targets, so a strong classical baseline is needed before any claim about a surrogate.
(iii)~The neural ensemble needs 32\% fewer new target evaluations, and in a separate source-only test its best first proposal per target beats that of the warm-start model. It takes 1.6 times the wall time of GP-BO here; whether its savings in evaluations become savings in time for costlier evaluations remains to be tested.
(iv)~The collisions conserve the magnitude of the bare Dicke overlap and the total-spin weights, an exact support ceiling holds on the line \(\gamin=\pi/2\), and for the best controls of a 12-target diagnostic panel most of the infidelity is an amplitude deficit that no final phase layer can remove.
(v)~Phase fitting produces large errors at a few isolated snapshots, but a stronger fitter leaves the ranking of the selected controls on that panel unchanged.

Section~\ref{sec:protocol} defines the protocol and the operational fidelity, and Sec.~\ref{sec:physics} derives the exact constraints. Section~\ref{sec:methods} describes the optimizers and how they are compared. The results are given in Sec.~\ref{sec:results} and discussed in Sec.~\ref{sec:discussion}, together with interleaved control, hardware cost, and the limitations of the study. Simulation conventions, optimizer settings, phase-fitting diagnostics, and reproducibility details are collected in the appendices.

\section{Collision protocol and fidelity objective}\label{sec:protocol}

\subsection{Dicke target and qubit arrangement}

For a register of \(N\) qubits with \(K\) excitations, the symmetric Dicke target is
\begin{equation}
\Dstate{N}{K}=\frac{1}{\sqrt{\Cnk}}
\sum_{\substack{z\in\{0,1\}^{N}\\ |z|=K}}\ket{z},
\label{eq:dicke}
\end{equation}
where \(|z|=\sum_{j=0}^{N-1}z_j\) is the Hamming weight of the bit string \(z\). The target is the normalized equal-weight superposition of all \(\Cnk\) configurations with \(K\) excitations; for \(K=1\) it is the \(W\) state.

The collision protocol starts from
\begin{equation}
\ket{\psi_0}=\ket{1}^{\otimes K}\otimes\ket{0}^{\otimes L},
\qquad L=N-K .
\label{eq:initial}
\end{equation}
The first \(K\) qubits are the excited shuttles. The remaining \(L\) qubits form two register chains, \(R\) and \(S\), of lengths \(r=\lfloor L/2\rfloor\) and \(s=L-r\). The shuttles are part of the final \(N\)-qubit state; no shuttle is measured, reset, postselected, or traced out.

A collision between sites \(i\) and \(j\) is described by
\begin{equation}
U_{ij}(\gamma)=e^{i\gamma\SWAP_{ij}}
             =\cos\gamma\,I+i\sin\gamma\,\SWAP_{ij},
\label{eq:pswap}
\end{equation}
where \(I\) is the two-qubit identity and \(\SWAP_{ij}\) exchanges the states of qubits \(i\) and \(j\). In terms of Pauli operators,
\begin{equation}
\SWAP_{ij}=\frac12
\left(I+X_iX_j+Y_iY_j+Z_iZ_j\right).
\label{eq:swap-pauli}
\end{equation}
Up to a global phase, Eq.~\eqref{eq:pswap} is an isotropic exchange interaction, and it conserves the total number of excitations. Its special values are \(U(0)=I\), \(U(\pi/2)=i\cdot\SWAP\), and \(U(\pi)=-I\), where \(i\cdot\SWAP\) is the SWAP operator times the global phase \(i\) and should be distinguished from the \textsc{iSWAP} gate. An angle close to \(\pi\) is thus a weak exchange with negative angle, up to a global phase, and is far from a full SWAP.

\subsection{Two-angle collision schedule}

The two control parameters are \(\gamin\), used for nearest-neighbor collisions inside either register, and \(\gamsh\), used for collisions between the active shuttle and a register qubit. We call them collision angles; through Eq.~\eqref{eq:pswap} they set the interaction strengths. In one shuttle subround, the active shuttle visits all \(r+s=L\) register sites in the fixed order of Ref.~\cite{vu2026pra}, and collisions between neighboring register sites are interleaved with these visits. Each register edge is acted on twice per subround, with other collisions in between, so the gates cannot be merged as if they commuted.

Subround \(t=1,2,\ldots\) uses shuttle
\begin{equation}
a_t=(t-1)\bmod K,
\label{eq:shuttle-selection}
\end{equation}
with qubits indexed from zero. One subround contains
\begin{equation}
m_{\mathrm{sh}}=L,\qquad
m_{\mathrm{in}}=2(r-1)+2(s-1)=2L-4
\label{eq:interaction-count}
\end{equation}
shuttle--register and intra-register collisions, respectively. These numbers count logical interactions of the protocol; the number of native gates on a given device is generally different. For \(\gamin=0\) the intra-register operations are identities and can be omitted.

We simulate \(T=200\) subrounds. As in Refs.~\cite{vu2026pra,nguyen2026mrz}, a round is one cycle through all \(K\) shuttles, that is, \(K\) consecutive subrounds, so \(T=200\) subrounds contain \(\lfloor200/K\rfloor\) complete rounds, followed by a partial round when \(K\) does not divide 200. For each candidate pair \(\boldsymbol\gamma=(\gamin,\gamsh)\) we compute the states
\begin{equation}
\ket{\psi_t(\boldsymbol\gamma)}
   =U_t(\boldsymbol\gamma)\cdots U_1(\boldsymbol\gamma)\ket{\psi_0},
\qquad t=0,\ldots,200,
\label{eq:evolution}
\end{equation}
where \(U_t\) is the ordered product of the gates of subround \(t\). Appendix~\ref{app:simulation} lists the register indices and the complete gate order, including the treatment of the register ends.

\begin{figure}[t]
\centering
\includegraphics[width=\columnwidth]{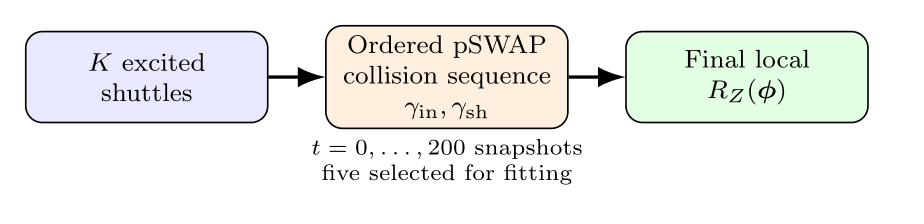}
\caption{Preparation and evaluation sequence (schematic, not a gate-level circuit). The excited shuttles (blue) collide with two initially empty registers through an ordered sequence of partial-SWAP (pSWAP) gates (orange) with fixed angles \(\gamin\) and \(\gamsh\). A final layer of local \(R_Z\) rotations (green) is fitted numerically at selected snapshots; in this work no phases are applied between collisions.}
\label{fig:scheme}
\end{figure}

\subsection{Raw, amplitude-only, and locally corrected fidelity}

Because the partial SWAP conserves the excitation number, each normalized snapshot can be written as
\begin{equation}
\ket{\psi_t}=\sum_{|z|=K}a_z(t)\ket z,\qquad
\sum_{|z|=K}|a_z(t)|^2=1.
\label{eq:sector-state}
\end{equation}
Its overlap with the Dicke target, which we call the raw fidelity, is
\begin{equation}
\Fin(t)=\bigl|\bra{D_N^{(K)}}\psi_t\rangle\bigr|^2
       =\frac{|\sum_{|z|=K}a_z(t)|^2}{\Cnk}.
\label{eq:raw}
\end{equation}
If the phase of every amplitude could be adjusted independently, the best fidelity attainable with the same amplitude magnitudes would be
\begin{equation}
\Fabs(t)=\frac{\left(\sum_{|z|=K}|a_z(t)|\right)^2}{\Cnk}.
\label{eq:abs}
\end{equation}
We call \(\Fabs\) the amplitude bound. It would be attained if every basis amplitude could be given its own phase; the single-qubit \(R_Z\) layer used below generally cannot do this, so \(\Fabs\) is an upper bound on the locally corrected fidelity at that snapshot.

The final correction is a product of single-qubit \(Z\) rotations,
\begin{equation}
V(\boldsymbol\phi)=\bigotimes_{j=0}^{N-1}
R_Z(\phi_j),\qquad
R_Z(\phi)=e^{-i\phi Z/2}.
\label{eq:rz}
\end{equation}
In a sector with fixed \(K\), the factor \(e^{-i\sum_j\phi_j/2}\) is a global phase, and the corrected fidelity is
\begin{equation}
F_t(\boldsymbol\phi)=
\frac1{\Cnk}\left|
\sum_{|z|=K} a_z(t)e^{i\sum_j z_j\phi_j}
\right|^2.
\label{eq:phase-objective}
\end{equation}
Adding the same constant to every \(\phi_j\) changes only the global phase, so Eq.~\eqref{eq:phase-objective} has \(N-1\) independent phase parameters.

The ideal objectives, maximized over the phase torus and over all snapshots, are
\begin{align}
F_t^\star&=\max_{\boldsymbol\phi\in\mathbb{T}^{N}}F_t(\boldsymbol\phi),\\
F_{\mathrm{ideal}}^{(200)}(\boldsymbol\gamma)
 &=\max_{0\le t\le200}F_t^\star.
\label{eq:ideal-score}
\end{align}
The numerical routine used in this work does not certify either maximum.

\subsection{Finite operational fidelity oracle}\label{subsec:oracle}

For each candidate \(\boldsymbol\gamma\), the oracle used in all benchmarks evolves the state for 200 subrounds and computes \(\Fabs(t)\) at the 201 snapshots \(t=0,\ldots,200\). It keeps the five snapshots with the largest amplitude bounds, \(\mathcal T_5(\boldsymbol\gamma)\), and fits local \(R_Z\) phases only at these five times. At each of them the phase vector is fitted with five Adam~\cite{kingma2014} restarts of 600 steps, with an initial learning rate of 0.05 and cosine decay. The first restart begins at the zero phase vector and the other four at Gaussian random phases with standard deviation 0.5~rad. The operational score is
\begin{equation}
\Fop(\boldsymbol\gamma;\omega,B)
=\max_{\substack{t\in\mathcal T_5\\
                 q=1,\ldots,5}}
 F_t\!\left(\boldsymbol\phi_{t,q}^{\mathrm{final}}\right),
\label{eq:operational-score}
\end{equation}
where \(\omega\) is the random-number state and \(B\) the batching convention, that is, how candidates are grouped when they are scored, which affects the random initial phases. The score is the maximum over the final iterates of the five restarts at each of the five selected snapshots, so it is a lower bound on the ideal objective:
\begin{equation}
\Fop\leq\max_{t\in\mathcal T_5}F_t^\star
\leq F_{\mathrm{ideal}}^{(200)}
\leq\max_{0\le t\le200}\Fabs(t)\leq1.
\label{eq:hierarchy}
\end{equation}
The inequality \(\Fin\leq\Fabs\) is exact. The inequality \(\Fop\geq\Fin\), on the other hand, is not guaranteed, because the Adam implementation does not keep the zero-phase starting value as a fallback. Throughout the paper we distinguish the ideal fidelity, its upper bounds, and the operational score.

\section{Exact constraints and physically relevant diagnostics}\label{sec:physics}

\subsection{Conservation of the bare Dicke projection}

Permutation symmetry gives the simplest exact property of the protocol. Since \(\SWAP_{ij}\Dstate{N}{K}=\Dstate{N}{K}\) for every pair of sites,
\begin{equation}
\bra{D_N^{(K)}}U_{ij}(\gamma)
 =e^{i\gamma}\bra{D_N^{(K)}}.
\label{eq:dicke-left}
\end{equation}
This relation can be applied gate by gate through the ordered sequence without any commutation assumption. With the interaction counts of Eq.~\eqref{eq:interaction-count} we obtain
\begin{align}
\bra{D_N^{(K)}}\psi_t\rangle
 &=e^{it[(2L-4)\gamin+L\gamsh]}
   \bra{D_N^{(K)}}\psi_0\rangle\nonumber\\
 &=\frac{e^{it[(2L-4)\gamin+L\gamsh]}}{\sqrt{\Cnk}},
\label{eq:raw-phase}
\end{align}
and hence
\begin{equation}
\Fin(t,\boldsymbol\gamma)=\frac{1}{\Cnk}.
\label{eq:raw-invariant}
\end{equation}

The bare overlap with the Dicke state thus has the same magnitude at every subround and for every pair of collision angles, although the collisions strongly redistribute populations and phases among the basis states. Large corrected fidelities are still possible because a product of local \(R_Z\) rotations with unequal angles is not permutation invariant. Equation~\eqref{eq:raw-invariant} constrains the raw overlap only and gives no bound on the corrected fidelity.

\subsection{Conserved total-spin-sector weights}

Collective angular momentum provides further invariants. Let \(\boldsymbol J=\tfrac12\sum_{j=0}^{N-1}(X_j,Y_j,Z_j)\). Qubit permutations commute with \(J_x,J_y,J_z\) and \(J^2=J_x^2+J_y^2+J_z^2\), so every partial-SWAP gate preserves the decomposition of the state into total-spin sectors. With \(J_z=\tfrac12\sum_j Z_j\) and \(Z\ket1=-\ket1\), the initial basis state has \(M=N/2-K\).

For \(K\leq N/2\) the allowed sectors are \(J=N/2-j\) with \(j=0,\ldots,K\). At fixed \(M\), the spin-\(J\) multiplicity space has dimension
\begin{equation}
d_{N,j}=\binom{N}{j}-\binom{N}{j-1},
\qquad \binom{N}{-1}=0.
\label{eq:spin-mult}
\end{equation}
Let \(P_J\) be the projector onto the subspace with given \(J\) and \(M\). It commutes with all qubit permutations, so its diagonal elements are equal on all \(\Cnk\) weight-\(K\) basis strings. Its trace over this subspace is \(d_{N,j}\), and the weight of the initial state in sector \(J=N/2-j\) is therefore
\begin{equation}
p_{J=N/2-j}
 =\bra{\psi_0}P_J\ket{\psi_0}
 =\frac{\binom Nj-\binom N{j-1}}{\Cnk}.
\label{eq:spinweights}
\end{equation}
These weights stay constant during the collision sequence. For the maximal spin \(J=N/2\), Eq.~\eqref{eq:spinweights} gives \(p_{N/2}=1/\Cnk\), in agreement with Eq.~\eqref{eq:raw-invariant}.

A layer of local \(Z\) rotations with unequal angles preserves the excitation number but in general does not commute with \(J^2\), so it can move weight between spin sectors. Since the maximal-spin sector is one dimensional at fixed \(M\), the final Dicke fidelity is the weight that the correction places in that sector. All collision angles start from the same conserved weights, however, and Eq.~\eqref{eq:spinweights} alone cannot tell which angles produce a state that local rotations can align well.

\subsection{Periods and reflection of the exact dynamics}

Since \(\SWAP^2=I\),
\begin{equation}
U(\gamma+\pi)=-U(\gamma),\qquad
U(\pi-\gamma)=-U(\gamma)^*.
\label{eq:gate-sym}
\end{equation}
The dynamics is periodic in each collision angle with period \(\pi\), up to a global phase. The dynamics also has a joint reflection symmetry, in which both angles are reflected at once,
\begin{equation}
\ket{\psi_t(\pi-\gamin,\pi-\gamsh)}
 =(-1)^{t(3L-4)}\ket{\psi_t(\gamin,\gamsh)}^*,
\label{eq:joint-reflection}
\end{equation}
where complex conjugation acts in the computational basis and the initial state is real. The reflected phases \(-\boldsymbol\phi\) give the same ideal corrected fidelity. Reflecting only one of the two angles is in general not a symmetry; we found explicit counterexamples at fixed subround.

We use these symmetries to check the simulator and to build a periodic covariance function for the Gaussian-process surrogate (Sec.~\ref{subsec:gp}); whether this kernel helps the search is examined separately in Sec.~\ref{subsec:res-transfer}. A finite, randomly initialized phase fit returns identical values for reflected angles only if, in addition, the initial phases are conjugated and the batch shapes are matched.

\subsection{Singular couplings and an exact support ceiling}\label{subsec:ceiling}

Special choices of the collision angles restrict the reachable states exactly~\cite{nguyen2026mrz}. For \(\gamsh\in\{0,\pi\}\), every shuttle--register collision is proportional to the identity, so the excitations stay on the shuttles, each snapshot is the initial basis string, and \(F_t(\boldsymbol\phi)=1/\Cnk\) for all \(t\) and \(\boldsymbol\phi\). The same happens at the center \((\gamin,\gamsh)=(\pi/2,\pi/2)\) of the search domain, where each collision is a permutation up to a phase. The search domains used here, \([0.05,\pi-0.05]^2\) for GP-BO and the classical searches and \([0.02,\pi-0.02]^2\) for the neural proposals (Appendix~\ref{app:search}), exclude the first case and contain the second.

A less obvious restriction appears on the line \(\gamin=\pi/2\), where every intra-register collision is a full SWAP. The analysis in Ref.~\cite{nguyen2026mrz} shows that for \(N-K\geq4\) the occupations of two register sites, tracked through these permutations, are conserved. Both sites are empty in \(\ket{\psi_0}\), so the state stays inside the \(\binom{N-2}{K}\) weight-\(K\) configurations that leave them empty, and the Cauchy--Schwarz inequality gives
\begin{equation}
F_t(\boldsymbol\phi)\leq\Fabs(t)\leq\frac{\binom{N-2}{K}}{\Cnk}
=\frac{(N-K)(N-K-1)}{N(N-1)}
\label{eq:support-ceiling}
\end{equation}
for every phase vector. The confinement holds after each collision and not only at the end of a round, so Eq.~\eqref{eq:support-ceiling} applies to all 201 snapshots scored by our oracle and to all targets considered here (\(N-K\geq6\)). Across the 65 targets the ceiling lies between 0.227 for \(\Dstate{12}{6}\) and 0.805 for \(\Dstate{20}{2}\), and for each of the 61 targets with a reference value it is below the best-found reference fidelity in Table~\ref{tab:reference-values}. Close to the line, the number of intra-register collisions needed to exceed the ceiling grows at least as \(|\gamin-\pi/2|^{-1}\)~\cite{nguyen2026mrz}. The optimization landscape therefore contains a line through the middle of the search domain on which the fidelity is capped exactly, surrounded by a region where exceeding the cap requires many collisions.

\subsection{Amplitude uniformity is a full-sector property}

With \(p_z=|a_z|^2\) the probability of configuration \(z\), the amplitude bound of a normalized state reads
\begin{equation}
\Fabs=\frac{\bigl(\sum_z\sqrt{p_z}\bigr)^2}{\Cnk}.
\label{eq:hellinger}
\end{equation}
It equals one exactly when all configurations have the same magnitude \(1/\sqrt{\Cnk}\). Writing \(\mathrm{CV}(|a|)\) for the coefficient of variation of the \(\Cnk\) amplitude magnitudes, computed with their arithmetic mean and variance, one has the identity
\begin{equation}
\Fabs=\frac{1}{1+\mathrm{CV}(|a|)^2}.
\label{eq:cv}
\end{equation}

One-qubit populations \(p_i=\langle n_i\rangle\), with \(n_i=(I-Z_i)/2\), do not determine the probability distribution over the weight-\(K\) configurations. The normalized state \((\ket{1100}+i\ket{0011})/\sqrt2\), for instance, has \(p_i=1/2\) on every site but \(\Fabs=1/3\) with respect to \(\Dstate{4}{2}\). Equal local populations therefore do not imply a uniform distribution over the full excitation sector. The sector entropy and the participation ratio describe this distribution better, but neither tells whether the relative phases can be corrected by local rotations.

\subsection{Many-body phase compatibility}

Perfect alignment of the nonzero sector amplitudes by local rotations requires
\begin{equation}
\arg a_z+\sum_jz_j\phi_j=\chi
\quad(\mathrm{mod}\ 2\pi)
\label{eq:phase-compatible}
\end{equation}
for every configuration \(z\) in the support, with one common phase \(\chi\). This is far more restrictive than giving each basis state its own correcting phase.

Four weight-\(K\) strings whose occupation vectors satisfy \(z_1+z_4=z_2+z_3\) provide a gauge-invariant test. Define
\begin{equation}
\Delta=\mathrm{Arg}
 \left(a_{z_1}a_{z_4}a_{z_2}^{*}a_{z_3}^{*}\right).
\label{eq:phase-plaquette}
\end{equation}
The phase \(z\cdot\boldsymbol\phi\) contributed by local rotations cancels in Eq.~\eqref{eq:phase-plaquette}, so a nonzero \(\Delta\) on nonvanishing amplitudes rules out perfect local alignment. A set of sampled plaquette phases \(\Delta\) alone does not fix the optimal fidelity or give a quantitative bound below \(\Fabs\), since the amplitude weights and the unsampled relations also enter.

For the snapshot chosen by the oracle we use the decomposition
\begin{equation}
1-\Fop=(1-\Fabs)+(\Fabs-\Fop).
\label{eq:gap-decomp}
\end{equation}
The first term is the amplitude deficit, which phase-only operations cannot remove at that state. The second combines phase incompatibility with the error of the finite numerical fit. Both refer to the selected snapshot only.

\subsection{Why a population-only recurrence is insufficient}\label{subsec:popwarm}

The exact population update for one collision with angle \(\gamma\) contains the two-site coherence \(\rho_{10,01}=\langle10|\rho|01\rangle\):
\begin{subequations}\label{eq:pop-update}
\begin{align}
p_i'&=\cos^2\gamma\,p_i+\sin^2\gamma\,p_j
 +2\cos\gamma\sin\gamma\,\mathrm{Im}\rho_{10,01},\\
p_j'&=\sin^2\gamma\,p_i+\cos^2\gamma\,p_j
 -2\cos\gamma\sin\gamma\,\mathrm{Im}\rho_{10,01},
\end{align}
\end{subequations}
where site \(i\) is the first qubit in the ordered basis \(\{\ket{10},\ket{01}\}\). The population-only warm-start model, which we use to propose initial angles, drops the coherence term, which need not stay small once correlations build up. Two successive collisions at \(\pi/4\) starting from \(\ket{10}\), for example, move the excitation completely to \(\ket{01}\) up to a phase, whereas the population-only recurrence predicts \(p_i=1/2\). This warm-start model is useful for proposing inexpensive starting angles, but it cannot describe the quantum state or predict the corrected fidelity.

\section{Numerical methods and comparison design}\label{sec:methods}

\subsection{Targets, empirical reference, and evaluation cost}

The study covers the 65 Dicke states with
\begin{equation}
11\leq N\leq20,\qquad
2\leq K\leq\left\lfloor\frac N2\right\rfloor.
\label{eq:target-family}
\end{equation}
The \(W\)-state case \(K=1\) is allowed by the protocol but is not part of this benchmark. To put the new searches in context, we built a composite reference from earlier, more extensive searches of the same protocol. Reference~\cite{vu2026pra} started from a dense grid of angle pairs, refined them with bound-constrained multistart L-BFGS-B~\cite{byrd1995}, and reported targets with up to 14 qubits; we keep its 31 numerical values for \(N=5\)--14, of which 16 fall in our range \(N=11\)--14. We added 45 values for \(N=15\)--20, obtained by extending the same angle search on a GPU. These 45 values were computed for the present comparison and are not part of Ref.~\cite{vu2026pra}. All reference values that overlap with our targets are listed in Table~\ref{tab:reference-values} (Appendix~\ref{app:repro}).

Sixty-one of the 65 targets have a reference value. For \(\Dstate{12}{6}\), \(\Dstate{13}{6}\), \(\Dstate{14}{6}\), and \(\Dstate{14}{7}\) none is available, and we leave these entries empty rather than fill them with results from a new optimizer. The 61 values average 0.771 (0.770529 to six decimals). All family means in our tables and figures are taken over these 61 targets, while every new method is run on all 65. The one exception is a contrast between the two GP variants whose definition and 59-target set were fixed before the full-family runs; it is described in Sec.~\ref{subsec:res-transfer}.

The reference shows the fidelity level reached by the earlier searches and is not a matched-budget benchmark. The earlier study used an evolution horizon of \(R=200K\) in its shuttle-sweep convention, whereas we stop after 200 subrounds, and its multistart search and phase fitting differ from the oracle of Sec.~\ref{subsec:oracle}. Differences between the reference mean and our means therefore cannot be attributed to the search algorithm alone.

A candidate evaluation is counted each time an angle pair \((\gamin,\gamsh)\) is submitted to the oracle of Eq.~\eqref{eq:operational-score}, which simulates 200 subrounds and performs the finite phase fit. A repeated coordinate is counted again whenever the oracle is rerun, and the internal Adam restarts are not counted as separate candidates. New target evaluations, observations reused from another method, and wall time measure different costs, and we report them separately.

\subsection{Neural surrogate across target systems}\label{subsec:neural}

The neural model describes a target and an angle pair by four dimensionless features,
\begin{equation}
\boldsymbol x(N,K,\boldsymbol\gamma)
=\left(\frac N{20},\,\frac KN,\,
        \frac{\gamin}{\pi},\,\frac{\gamsh}{\pi}\right).
\label{eq:features}
\end{equation}
An ensemble~\cite{lakshminarayanan2017} of five networks (three residual multilayer perceptrons and two mixture-of-experts models~\cite{jacobs1991}) predicts the operational fidelity. Each network is trained with the fidelity-weighted loss
\begin{equation}
\mathcal L_{\mathrm{sur}}=
\frac1{M_{\mathrm{train}}}
\sum_{i=1}^{M_{\mathrm{train}}}(1+4F_i)
  \bigl[\widehat F(\boldsymbol x_i)-F_i\bigr]^2,
\label{eq:nnloss}
\end{equation}
where \(M_{\mathrm{train}}\) is the number of training observations and \(F_i\) the oracle score. The factor \(1+4F_i\) gives more weight to high-fidelity examples. The ensemble learns the operational score from data; unlike physics-informed approaches~\cite{norambuena2024}, it does not use the Schr\"odinger equation as a training residual.

The ensemble mean and member dispersion are
\begin{equation}
\mu(\boldsymbol x)=\frac15\sum_{m=1}^{5}\widehat F_m(\boldsymbol x),
\qquad
s^2(\boldsymbol x)=\frac15\sum_{m=1}^{5}
  [\widehat F_m(\boldsymbol x)-\mu(\boldsymbol x)]^2.
\label{eq:ensemble}
\end{equation}
Both enter the selection of candidates, although the dispersion is a heuristic spread rather than a calibrated uncertainty. New angle pairs are chosen by gradient-based proposal searches with a fixed boundary penalty and a diversity rule across ensemble members. Their oracle scores are added to a training buffer shared by all larger targets, and targets whose best fidelity is still low receive more proposals.

Before any larger target is optimized, the ensemble is trained on 750 oracle observations from the 15 Dicke states with \(N=5\)--10 and \(2\le K\le\lfloor N/2\rfloor\). Each of these small targets contributes 25 Latin-hypercube angle pairs and 25 analytical warm-start proposals. The 750 training labels were generated for this study and are separate from the composite reference. The same source buffer is reused in five complete searches of the larger targets, with optimizer seeds \(2026,9,11,919,20961\); only the 750 completed evaluations were kept, without the full record of attempts and retries. Appendix~\ref{app:search} gives the architecture, the training and acquisition schedule, and the initialization.

\subsection{Gaussian-process and classical alternatives}\label{subsec:gp}

Standard GP-BO~\cite{snoek2012} fits a separate model for each \((N,K)\). Its first ten angle pairs are five target-dependent proposals from the population-only warm-start model of Sec.~\ref{subsec:popwarm} and five scrambled Sobol points, all scored by the full oracle before model-based acquisition starts. Eight further batches of five candidates are then chosen with the numerically stable batch logarithmic expected improvement (qLogEI)~\cite{ament2023} implemented in BoTorch~\cite{balandat2020}, for a total of 50 evaluations per target. The covariance is a Mat\'ern-\(5/2\) kernel~\cite{rasmussen2006},
\begin{equation}
k(r)=\sigma_f^2\left(1+\sqrt5\,r+\frac53r^2\right)e^{-\sqrt5r},
\label{eq:matern}
\end{equation}
where \(r\) is a distance in angle space and \(\sigma_f^2\) the fitted output scale. On the domain
\begin{align}
\gamin,\gamsh&\in[0.05,\pi-0.05],\nonumber\\
u_a&=(\gamma_a-0.05)/w,\qquad w=\pi-0.10,
\label{eq:gp-domain}
\end{align}
the standard kernel uses learned length scales \(\ell_a\) through
\begin{equation}
d_A^2(\boldsymbol\gamma,\boldsymbol\gamma')
=\sum_{a\in\{\mathrm{in},\mathrm{sh}\}}
 \frac{(\gamma_a-\gamma'_a)^2}{(w\ell_a)^2}.
\label{eq:ordinary-distance}
\end{equation}
A second variant builds the exact \(\pi\) periodicity of each angle into the distance,
\begin{equation}
d_B^2(\boldsymbol\gamma,\boldsymbol\gamma')
=\sum_a\frac{4\sin^2(\gamma_a-\gamma_a')}{\ell_a^2}.
\label{eq:periodic-distance}
\end{equation}
With the same numerical length-scale priors the two distances are not matched at short range, so a performance difference between the variants cannot be attributed to the periodic geometry alone. The three-seed GP study uses optimizer seeds \(211,503,1009\), the same 65 targets, and 50 evaluations per target for each variant.

We also compare with searches that use no surrogate model. The nonadaptive controls are a Sobol continuation, which adds 40 new Sobol points to the ten initial GP observations, and a fixed 50-point warm-start grid with its own initial design. The adaptive classical controls are bounded Nelder--Mead~\cite{nelder1965}, Powell's method~\cite{powell1964}, and a two-basin pattern search. Each classical control reuses the ten initial GP observations and spends 40 new evaluations. Nelder--Mead and Powell submit one candidate at a time, whereas GP-BO and the pattern search normally submit batches of five. The random restarts of the phase fit depend on the batch shape and on the random-number streams, so equal numbers of candidates do not give identical stochastic realizations of the oracle. Appendix~\ref{app:search} gives the basin initialization and the acquisition settings.

\subsection{Transfer tests and physical diagnostics}

Two separate experiments test the reuse of smaller-system data. In the source-only experiment, a neural ensemble trained only on the 750 source observations proposes five angle pairs for each larger target before any of its fidelities is known. In the source-ablation experiment, we compare networks trained with and without the 750 source observations while both also learn from larger-target evaluations, starting from two, five, or ten initial observations per target and followed by shared active updates. The two experiments address different aspects of transfer.

Finally, we study the dynamics and the oracle at fixed controls, without new searches, for twelve diagnostic targets,
\begin{align*}
(N,K)\in\{&
(11,2),(11,5),(12,6),(14,5),\\
&(15,7),(16,2),(16,8),(18,8),\\
&(19,5),(20,2),(20,8),(20,10)
\}.
\end{align*}
For each target we take five angle pairs: the best GP-BO, neural, and Nelder--Mead controls, one analytical warm-start control chosen without evaluating its fidelity, and a perturbed copy of the GP angles. One pair appears twice, so the 60 labeled controls give 59 distinct trajectories. Each is followed from \(t=0\) to \(200\), with scalar diagnostics at every subround. A follow-up study stores complex amplitudes at selected snapshots and compares the original Adam fit with monotone coordinate ascent and an independent L-BFGS fit~\cite{byrd1995}. These diagnostics test the oracle and leave the controls and scores of the optimization tables unchanged. Appendix~\ref{app:phase} gives the details.

\section{Results}\label{sec:results}

\subsection{Adaptive sampling versus fixed designs and local search}\label{subsec:res-adaptive}

Table~\ref{tab:methods-50} gives the mean best-found operational fidelity at the end of each search. For each method we average the optimizer seeds within a target and then average over the 61 targets with equal weight. The five-seed neural/GP comparison and the three-seed GP/classical comparison are separate experiments, and rows from different studies should not be read as paired results. For comparison, the composite reference averages 0.771 over the same targets.

\begin{table*}[t]
\caption{Mean best-found operational fidelity \(\Fop\) at the end of each search, averaged over the 61 targets with a reference value. All methods use the oracle of Sec.~\ref{subsec:oracle}. GP-BO and the classical searches use 50 logical evaluations per target and seed, some of them reused initial observations; the neural search allocates new evaluations adaptively (34.0 per target, averaged over all 65 targets) and also uses a shared buffer of 750 smaller-system observations (Sec.~\ref{subsec:res-cost}). \(\Delta\) is the difference from standard GP-BO in the same study. The seed-level column gives the number of whole-family seeds in which the named method is ahead and, where available, the sample standard deviation (SD) of the paired seed differences.}
\label{tab:methods-50}
\centering
\newcommand{\rS}[1]{\parbox[t]{.10\textwidth}{\raggedright\strut #1\strut}}
\newcommand{\rM}[1]{\parbox[t]{.20\textwidth}{\raggedright\strut #1\strut}}
\newcommand{\rN}[1]{\parbox[t]{.065\textwidth}{\raggedleft\strut #1\strut}}
\newcommand{\rE}[1]{\parbox[t]{.25\textwidth}{\raggedright\strut #1\strut}}
\newcommand{\rC}[1]{\parbox[t]{.21\textwidth}{\raggedright\strut #1\strut}}
\begin{tabular}{@{}llllll@{}}
\toprule
\rS{Study} & \rM{Method} & \rN{Mean \(\Fop\)} & \rN{\(\Delta\)} & \rE{Seed-level evidence} & \rC{Evaluations per target and seed}\\
\midrule
\rS{Five seeds} & \rM{Neural surrogate} & \rN{0.766} & \rN{\(+0.002\)} & \rE{Neural ahead in 4 of 5 seeds; SD 0.003} & \rC{34.0 new (mean, adaptive); 750-observation source buffer}\\
\rS{} & \rM{Standard GP-BO} & \rN{0.763} & \rN{---} & \rE{} & \rC{50 new}\\
\midrule
\rS{Three seeds} & \rM{Periodic GP-BO} & \rN{0.767} & \rN{\(+0.004\)} & \rE{} & \rC{50 new}\\
\rS{} & \rM{Standard GP-BO} & \rN{0.762} & \rN{---} & \rE{} & \rC{50 new}\\
\rS{} & \rM{Bounded Nelder--Mead} & \rN{0.760} & \rN{\(-0.002\)} & \rE{GP ahead in 3 of 3 seeds (SD 0.003); Nelder--Mead has the higher three-seed mean on 41 of 61 targets} & \rC{40 new + 10 reused}\\
\rS{} & \rM{Two-basin pattern search} & \rN{0.752} & \rN{\(-0.010\)} & \rE{} & \rC{40 new + 10 reused}\\
\rS{} & \rM{Bounded Powell} & \rN{0.621} & \rN{\(-0.142\)} & \rE{} & \rC{40 new + 10 reused}\\
\rS{} & \rM{Fixed warm-start grid} & \rN{0.699} & \rN{\(-0.063\)} & \rE{GP ahead in 3 of 3 seeds} & \rC{50 new, own initialization}\\
\rS{} & \rM{Fixed Sobol continuation} & \rN{0.674} & \rN{\(-0.088\)} & \rE{GP ahead in 3 of 3 seeds} & \rC{40 new + 10 reused}\\
\bottomrule
\end{tabular}
\end{table*}

In the three-seed study, GP-BO reaches 0.762. It is ahead of the Sobol continuation by 0.088 and of the warm-start grid by 0.063, in each of the three seeds. Adaptive sampling thus outperforms these two fixed designs at this budget; better nonadaptive designs in two dimensions may exist, but we did not search for them.

Bounded Nelder--Mead comes much closer. With the same budget it reaches 0.760, 0.002 below GP-BO. GP-BO has the higher family-wide mean in all three seeds (paired differences \(+0.001\), \(+0.005\), and \(+0.001\)), but the median of the target-level differences is \(-0.002\), and Nelder--Mead has the higher three-seed mean on 41 of the 61 targets (Fig.~\ref{fig:gp-nm-targets} in Appendix~\ref{app:search}). Most of the GP-minus-Nelder--Mead gap in the family mean comes from one outlier, \(\Dstate{19}{5}\), where GP-BO is ahead by 0.149; without this target, the mean difference over the other 60 is \(-5.5\times10^{-5}\). The family means of the two methods are thus nearly identical at this budget, and at the level of individual targets Nelder--Mead is ahead slightly more often. The two-basin pattern search ends within 0.010 of GP-BO, while our Powell configuration falls far behind. For a two-parameter problem of this kind, a surrogate must be compared with a well-configured local search before any advantage can be claimed.

\subsection{Budget dependence and the cost of learning across system sizes}\label{subsec:res-cost}

\begin{figure*}[t]
\centering
\includegraphics[width=0.9\textwidth]{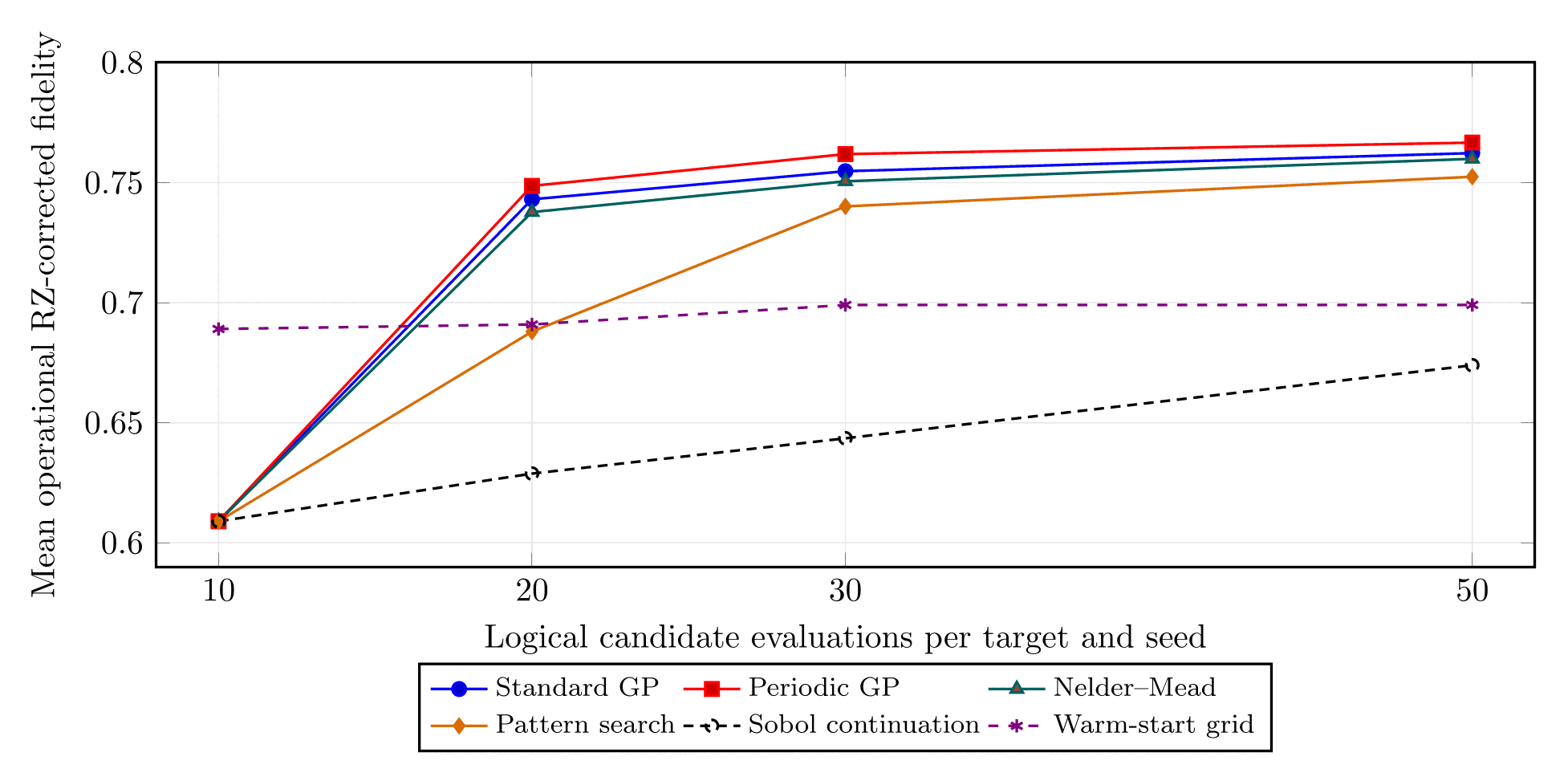}
\caption{Mean best-found operational fidelity versus the number of logical evaluations per target, for the 61 targets with a reference value, averaged over seeds 211, 503, and 1009. Points are at 10, 20, 30, and 50 evaluations. All curves except the warm-start grid (purple, dashed) begin from the same ten scored observations; the grid has its own predefined initialization. Nelder--Mead submits one candidate at a time, while GP-BO, pattern search, and the Sobol continuation submit batches of five.}
\label{fig:budget-curves}
\end{figure*}

Figure~\ref{fig:budget-curves} shows how the best-found mean grows with the budget in the three-seed study. With ten observations the fixed grid is ahead of GP-BO (0.689 against 0.609), because the two methods start from different designs. After 20 observations GP-BO has reached 0.743 and the grid only 0.691. The value of adaptive proposals depends on both the budget and the starting points.

In the five-seed study the neural surrogate reaches 0.766 and GP-BO 0.763 on the same 61 targets. The paired difference is \(+0.002\) (seed SD 0.003), and the neural method is ahead in four of the five seeds. It uses 11,049 new evaluations of the larger targets over five 65-target runs, against 16,250 for GP-BO, that is, 34.0 instead of 50 new evaluations per target. This 32\% reduction is made possible by the 750 smaller-system observations that the network has already seen.

The 750 source observations must also be counted. Charging all of them to a single 65-target run gives
\begin{equation}
\overline Q_{\rm NN}^{(1)}
=\frac{2209.8+750}{65}\simeq45.5
\label{eq:neural-amortized-cost}
\end{equation}
observations per target, where 2209.8 is the mean number of new evaluations in one neural run. Amortized over the five runs reported here, the cost is \((11049+750)/(5\times65)\simeq36.3\) observations per target and seed. Neither number includes network training or the proposal searches.

The wall time shows the other side of the trade-off. A complete 65-target run takes 94.8~min with the neural method and 58.5~min with GP-BO, where the neural time includes model training and proposal generation (Table~\ref{tab:five-seeds} in Appendix~\ref{app:search}). Averaged over these runs, GP-BO spends about 1.1~s per evaluation, including its modeling overhead, so evaluations are cheap in this problem. The neural method saves evaluations at the price of wall time. The balance may turn in its favor when a single evaluation is much more expensive, for example for larger registers or for evaluations on hardware; we have not measured where this crossover lies.

\subsection{Periodic covariance and transfer from smaller systems}\label{subsec:res-transfer}

In the three-seed study, the periodic GP variant reaches 0.767 at 50 evaluations, compared with 0.762 for standard GP-BO. Whether this difference reflects the periodic kernel was assessed with a primary contrast that we had defined before the full-family runs: the difference between the two variants after 20 calls, averaged over the 59 targets that were not part of the six-target pilot (Appendix~\ref{app:search}). These 59 targets include the four without a reference value, so this set differs from the 61-target set used elsewhere.

The contrast is \(+0.004\). Its descriptive target/seed bootstrap interval, \([-0.005,0.013]\), includes zero, and removing the most favorable target brings the contrast below the improvement margin we had set in advance for adopting the periodic kernel. A further confound is that the two kernels have different effective length scales under the same priors; a 12-target scale-control study did not identify the periodic geometry as the source of the difference. We therefore find no demonstrated advantage of the periodic kernel and keep it only as a physically motivated option.

Smaller-system data are most useful before any target data exist. Ensembles trained only on the 750 source observations propose five angle pairs per larger target. The best of the five averages 0.668, against 0.598 for five proposals from the population-only warm-start model, a gain of 0.070; the neural proposals are still worse on 25 of the 61 targets.

Once the network also learns from larger-target data, the effect of the source observations is unclear. With ten initial observations per target, the 50-call means are 0.763 with source data and 0.759 without, but the sign of the effect changes between seeds. With two initial observations, the advantage after ten calls is \(+0.002\), with seed differences of \(-0.002\), \(+0.028\), and \(-0.020\), and the five-observation variant is also mixed. The source data thus improve the initial proposals relative to the warm-start model, but we find no consistent benefit during the later active learning.

\subsection{Physical limits and oracle reliability}

The exact results of Sec.~\ref{sec:physics} help interpret the similar best-found fidelities of the leading adaptive searches. A final layer of local phases cannot change the amplitude magnitudes, so a high corrected fidelity needs a nearly uniform distribution over the entire weight-\(K\) sector; uniform one-body populations are not enough. We characterize the sector distribution by the Shannon entropy \(H=-\sum_{|z|=K}p_z\log p_z\), with \(p_z=|a_z|^2\) and \(0\log0=0\). The normalized entropy \(H/\log\Cnk\) equals one for a uniform distribution.

\begin{figure*}[t]
\centering
\includegraphics[width=0.9\textwidth]{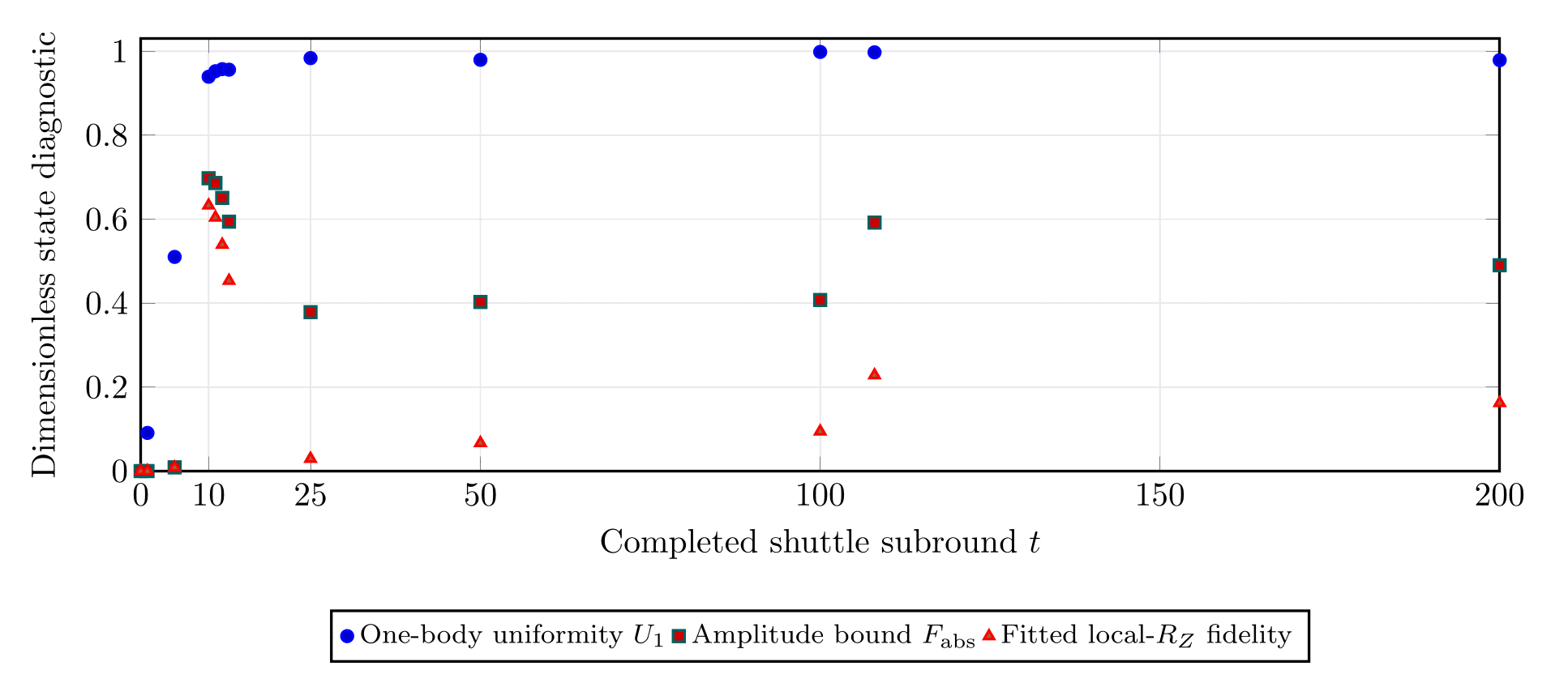}
\caption{Diagnostics at twelve subrounds (\(t=0,1,5,10,11,12,13,25,50,100,108,200\)) for the same selected GP-BO angle pair for \(\Dstate{20}{10}\). Circles: one-body uniformity \(U_1(t)=1-\operatorname{Var}_j[p_j(t)]/[f(1-f)]\), where \(p_j(t)\) is the occupation of qubit \(j\) and \(f=K/N=1/2\). Squares: amplitude bound \(\Fabs(t)\). Triangles: fidelity after fitting the final \(R_Z\) layer. The markers are not connected because the phases were fitted only at these snapshots. From \(t=10\) to \(100\) the populations become more uniform, while the amplitude bound and the corrected fidelity both decrease.}
\label{fig:mixing-timepoints}
\end{figure*}

A selected GP-BO angle pair for \(\Dstate{20}{10}\) illustrates the difference (Table~\ref{tab:mixing-counterexample} and Fig.~\ref{fig:mixing-timepoints}). Between subrounds 10 and 100 the one-qubit occupations become much more uniform, yet the amplitude bound and the fitted corrected fidelity both decrease strongly. Single-site populations fix only the marginals of the weight-\(K\) distribution; here the marginals approach their uniform values while the full distribution becomes less uniform, as the drop in sector entropy in Table~\ref{tab:mixing-counterexample} shows.

\begin{table}[t]
\caption{Two snapshots for a single selected GP-BO angle pair for \(\Dstate{20}{10}\), held fixed over time. The relative occupation spread is the standard deviation of the one-qubit populations divided by \(\sqrt{f(1-f)}\), with \(f=K/N\). Local phases were fitted only at the listed snapshots.}
\label{tab:mixing-counterexample}
\centering
\begin{tabular}{@{}lrr@{}}
\toprule
Observable & \(t=10\) & \(t=100\)\\
\midrule
Relative occupation spread & 0.246 & 0.039\\
Amplitude bound \(\Fabs(t)\) & 0.698 & 0.408\\
Fitted corrected \(F_t\) & 0.633 & 0.095\\
Sector entropy \(H/\log \Cnk\) & 0.903 & 0.788\\
\bottomrule
\end{tabular}
\end{table}

For the 36 selected GP-BO, neural, and Nelder--Mead controls of the 12-target diagnostic panel, the mean amplitude deficit \(1-\Fabs\) at the selected snapshot is 0.186, and the mean gap \(\Fabs-\Fop\), which combines phase incompatibility and fitting error, is 0.049 [Eq.~\eqref{eq:gap-decomp}]. The mean amplitude deficit thus accounts for 79.1\% of the mean infidelity of these selected states. Removing this part of the infidelity requires a different amplitude distribution, which a final phase layer cannot produce; it would have to come from other collision angles or from phases applied during the evolution (Sec.~\ref{subsec:interleaved}). Plaquette-phase statistics and stopping times are reported in Appendix~\ref{app:phase}.

The oracle itself can return very low values at individual snapshots. For \(\Dstate{14}{5}\) at subround 51, the original Adam fit gives 0.002. The phase vector fitted one subround earlier gives 0.843 on the same state, and monotone coordinate ascent reaches 0.854, close to the amplitude bound of 0.865 (Table~\ref{tab:phase-witness}). Independent rescoring confirms that this is a failure of the fit and not a different physical state.

\begin{table}[t]
\caption{Failure of the original Adam phase fit for \(\Dstate{14}{5}\) with the selected GP-BO angles at subround \(t=51\). All entries except the amplitude bound are fidelities attained on the same state. The best score of this control, 0.853 at \(t=50\), is not affected.}
\label{tab:phase-witness}
\centering
\begin{tabular}{@{}lr@{}}
\toprule
Phase selection & Fidelity at \(t=51\)\\
\midrule
Original five-restart Adam & 0.002\\
Previously fitted \(t=50\) phases & 0.843\\
Monotone coordinate ascent & 0.854\\
Independent L-BFGS & 0.854\\
Amplitude upper bound \(\Fabs\) & 0.865\\
\bottomrule
\end{tabular}
\end{table}

A stronger phase-fitting study reproduces all 12 failures of this kind that we had identified, and coordinate ascent improves 21 of 744 snapshot fits by more than 0.01. For the 36 selected controls of the 12-target panel, however, the final scores change by only \(6.9\times10^{-5}\) on average (maximum 0.0011), and the ranking of the GP-BO, neural, and Nelder--Mead controls stays the same for all 12 targets. Neither coordinate ascent nor L-BFGS certifies the global phase optimum; in six of 16 cross-checks L-BFGS ends up to 0.038 below coordinate ascent (Appendix~\ref{app:phase}). The reported scores are values of a well-defined finite oracle, and on the tested panel its ranking of the selected controls is stable.

\section{Discussion}\label{sec:discussion}

For the passive two-angle protocol, these results have a simple practical reading. For the relatively inexpensive evaluations of this implementation (about 1.1~s of wall time per GP-BO candidate, including modeling overhead), an adaptive local search performs about as well as a surrogate model: in the three-seed comparison, standard GP-BO exceeds bounded Nelder--Mead by only 0.002 in family-mean best-found fidelity. Reusing data across system sizes changes which resource is scarce: it lowers the number of new target evaluations but increases wall time. This agrees with comparisons of learning-based and conventional quantum control, in which the benefit of learning depends on the control constraints and the problem size~\cite{zhang2019rl}, and with benchmarks in which the ranking of derivative-free solvers depends on dimension and budget~\cite{rios2013}. Surrogates are more likely to pay off when evaluations become substantially more expensive, when the control schedule has more parameters, or when evaluations are noisy, as on hardware; none of these regimes was tested here.

The leading adaptive methods (standard and periodic GP-BO, the neural surrogate, and bounded Nelder--Mead) reach family-mean best-found fidelities of 0.760--0.767, close to the composite-reference average of 0.771. The exact results of Sec.~\ref{sec:physics} and the diagnostics of Sec.~\ref{sec:results} help explain this. The collisions conserve the magnitude of the bare Dicke overlap and the total-spin weights, the fidelity is capped exactly on the line \(\gamin=\pi/2\), and in the diagnostic panel most of the remaining infidelity lies in amplitude magnitudes that a final phase layer cannot change. That several different optimizers reach a similar level suggests that further gains need additional control freedom, although we have not optimized the full landscape of angles and stopping times globally.

\subsection{Minimal versus full phase control}\label{subsec:interleaved}

The passive protocol is the minimal-control member of a larger family. It implements \(V_Z(\boldsymbol\phi)\Ucoll(\boldsymbol\gamma)\), with all phases applied at the end. An interleaved protocol applies phases throughout,
\begin{equation}
U_{\mathrm{interleaved}}=
V_Z^{(T_{\mathrm r})}W_{T_{\mathrm r}}\,V_Z^{(T_{\mathrm r}-1)}W_{T_{\mathrm r}-1}\cdots
V_Z^{(1)}W_1,
\label{eq:interleaved}
\end{equation}
where \(W_k\) is the ordered product of the \(K\) subround unitaries of round \(k\), \(V_Z^{(k)}\) is a layer of single-qubit \(Z\) rotations, and \(T_{\mathrm r}\) is the number of rounds. Phases applied before later collisions change the subsequent interference, the amplitude magnitudes, and the total-spin weights, which a final layer cannot do (Sec.~\ref{sec:physics}). Interleaved phases can therefore act on the amplitude deficit identified above, the part of the error that a final layer leaves untouched.

Reference~\cite{nguyen2026mrz} studies this full-control family. With fixed collision angles and \(T_{\mathrm r}(N-1)\) independent phases over \(T_{\mathrm r}\) rounds, generic couplings give access to the full projective state space of the fixed-excitation sector in the systems tested there, and higher Dicke fidelities are reached for registers of up to 13 qubits. The ceiling of Eq.~\eqref{eq:support-ceiling} holds for both protocols, because diagonal phase layers preserve the invariant coordinate sectors. The price of full control is individual phase control of every qubit between rounds and an optimization over hundreds of phases instead of two collision angles.

Which protocol is preferable depends on the hardware. On superconducting processors, virtual \(Z\) gates add little or no gate time~\cite{mckay2017}, provided the native exchange gate can absorb the accumulated frame changes, and full phase control is then likely preferable. Where single-qubit control during the sequence is slow, error-prone, or unavailable, or where only a few global parameters can be calibrated or learned online, the passive protocol and the sample-efficient searches studied here are relevant options.

\subsection{Collision depth and hardware cost}

Our oracle looks for the best snapshot within 200 subrounds. In Ref.~\cite{vu2026data}, a single round already gave the highest noiseless fidelity found within the same depth range for 14 of 31 targets with \(N\leq14\). In that study, the deep-search operating points needed a median of 72.9 times the two-qubit gate count of one round with \(\gamin=0\), and on a superconducting processor the deep circuits kept a much smaller fraction of their noiseless fidelity than one-round circuits. In our diagnostic panel, 7 of the 12 selected GP-BO controls likewise select their best snapshot at \(t=K\), after one round (Appendix~\ref{app:phase}). An objective that penalizes depth would favor such shallow operating points and is a natural next target for the surrogate methods compared here.

\subsection{Scope and limitations}\label{subsec:scope}

All fidelities reported here are noiseless, finite-precision state-vector results. They are not hardware measurements and include neither decoherence nor compilation to native gates; imperfections of the same protocol are analyzed in Ref.~\cite{vu2026pra}. Each value is a best-found score of the oracle of Eq.~\eqref{eq:operational-score}, which fits phases only at the five snapshots with the largest amplitude bounds. It is a lower bound on the ideal objective of Eq.~\eqref{eq:ideal-score}, and none of our values certifies the global optimum over angles, stopping times, and phases. The composite reference comes from earlier searches with a different horizon and phase-fitting rule and without a record of attempts, so it indicates a level of achieved fidelity rather than serving as a matched-cost benchmark or a baseline for speedup claims. Statements about seeds rest on three or five whole-family repetitions, and the bootstrap intervals are descriptive. Wall times depend on our implementation and hardware. Oracle reliability was examined on a fixed 12-target panel, and our statements about ranking stability refer to that panel. Finally, all searches concern one two-parameter architecture. Whether the relative performance of the methods carries over to larger registers, noisy evaluations, or control schedules with more parameters remains to be tested.

\section{Conclusions}\label{sec:conclusions}

We studied methods for choosing the two collision angles of measurement-free, partial-SWAP Dicke-state preparation with a limited number of many-body evaluations, for 65 targets with 11--20 qubits. Adaptive sampling beats the tested fixed designs, and a well-configured Nelder--Mead search reaches nearly the same family-mean fidelity as Gaussian-process Bayesian optimization at the tested budget, so strong classical baselines are essential when surrogates are judged in low dimension. A neural surrogate that reuses data from smaller systems needs about a third fewer new target evaluations and, in a separate source-only test, gives better initial proposals than the warm-start model, although it needs more wall time in the present implementation; whether it saves total time for costlier evaluations remains to be tested. Exact conservation laws, a support ceiling on one line of the parameter plane, and an amplitude-dominated infidelity in the diagnostic panel help explain the similar family-mean fidelities of the leading adaptive methods, and they identify a limitation that interleaved phase control, the subject of Ref.~\cite{nguyen2026mrz}, can address. Independent phase-fitting checks show that the finite oracle can fail at isolated snapshots without changing the ranking of the selected controls on the diagnostic panel.

\begin{acknowledgments}
\"O.E.M. and F.O. acknowledge financial support from the Scientific and Technological Research Council of T\"urkiye (T\"UB\.ITAK), Project Grant No.~126F097. F.O. acknowledges financial support from the Tokyo International University Personal Research Fund and Special Grant-in-Aid for Research Work. AI-based tools assisted with drafting, editing, and checking the manuscript; the authors reviewed all content, numbers, and references and take full responsibility for the manuscript.
\end{acknowledgments}

\section*{Data availability}

The search software will be made available at \url{https://github.com/kayv297/surrogate}. The reference values are listed in Table~\ref{tab:reference-values} and the per-seed neural and GP results in Table~\ref{tab:five-seeds}. The run journals of the 155,423 recorded evaluations, with candidate angles, selected stopping subrounds, and fitted phases, and the phase-fitting archives are retained by the authors and will be deposited together with a tagged version of the code in a versioned public archive. The complex state trajectories of these evaluations were not stored, so their local phases cannot be refitted retrospectively; the phase-certification study kept the complex Dicke-sector amplitudes of 921 selected snapshots (Appendix~\ref{app:phase}).

\appendix
\section{Simulation and fidelity conventions}\label{app:simulation}

\subsection{Exact qubit indices and collision counts}

The simulation retains all \(N\) qubits in a dense state vector. Qubit indices \(0,\ldots,K-1\) label the initially excited shuttles. Sites \(K,\ldots,K+r-1\) form register \(R\), while \(K+r,\ldots,N-1\) form register \(S\), where \(r=\lfloor(N-K)/2\rfloor\) and \(s=N-K-r\). The initial ket is the computational basis string consisting of \(K\) ones followed by \(N-K\) zeros.

At subround \(t\), the active shuttle has index \(a_t=(t-1)\bmod K\). Denote the ordered register sites by \(R_j=K+j\), \(j=0,\ldots,r-1\), and \(S_j=K+r+j\), \(j=0,\ldots,s-1\). The evolution is specified by the following sequence of successive, generally noncommuting partial-SWAP gates:

\begin{enumerate}
\item For \(j=0,1,\ldots,\max(r,s)-1\), visit the two registers in the order \(R\) then \(S\).
\item If \(j<r\), first apply \(U_{a_t,R_j}(\gamsh)\). Next apply \(U_{R_j,R_{j-1}}(\gamin)\) if \(j>0\), followed by \(U_{R_j,R_{j+1}}(\gamin)\) if \(j<r-1\).
\item If \(j<s\), after completing the operations on \(R_j\), apply \(U_{a_t,S_j}(\gamsh)\). Next apply \(U_{S_j,S_{j-1}}(\gamin)\) if \(j>0\), followed by \(U_{S_j,S_{j+1}}(\gamin)\) if \(j<s-1\).
\item Continue to \(j+1\) without reordering any of the preceding gates. If \(\gamin=0\), the intra-register gates are identities and can be omitted.
\end{enumerate}

Each register qubit thus interacts once with the active shuttle per subround, and every nearest-neighbor register pair undergoes two collisions per subround. Other collisions occur between the two visits, so they cannot be merged into a single effective interaction. The initial state at \(t=0\) and the state after each individual shuttle subround constitute the 201 snapshots computed for scoring.

For the studied targets \(r,s\geq1\), the logical interaction counts for one subround are \(N-K\) shuttle and \(2(N-K)-4\) intra-register operations. At time \(t\), the cumulative logical interaction count is \(t[3(N-K)-4]\), including identity interactions at zero angle. It differs from the number of physical entangling gates after compilation; the simulations impose no hardware connectivity, routing, decoherence, or native two-qubit decomposition.

The maximum simulated time is \(T=200\) individual shuttle subrounds, yielding 201 computed snapshots including \(t=0\). The 200 subrounds contain \(\lfloor200/K\rfloor\) complete rounds, followed by a partial round when \(K\) does not divide 200. The analytical warm-start generator is separate from the quantum simulator. It propagates approximate one-qubit populations through 80 complete rounds (\(80K\) subrounds) to suggest initial angles, whereas every candidate is scored by the 200-subround state-vector calculation. Because the population recurrence neglects coherence [Eq.~\eqref{eq:pop-update}], it is used only to propose starting points.

\subsection{Numerical state and phase scorer}

State evolution uses the benchmark implementation with complex64/float32 arithmetic on an NVIDIA A10G GPU. Fidelity scoring extracts the amplitudes \(a_z\) in the fixed-\(K\) subspace. The returned score uses the state amplitudes as computed, without a hidden state renormalization step. Numerical norm drift and excitation leakage were audited separately. The physical-dynamics panel found maximum norm drift \(2.51\times10^{-4}\), maximum excitation-sector leakage \(4.77\times10^{-7}\), and maximum absolute discrepancy \(1.48\times10^{-6}\) from Eq.~\eqref{eq:raw-invariant}. These deviations are due to finite numerical precision.

The original benchmark phase routine computes \(\Fabs\) for the 201 time snapshots \(t=0,\ldots,200\), ranks those amplitude-only bounds, retains the five largest, and performs full local-RZ phase fitting only at these five times. The remaining 196 snapshots are not independently RZ-optimized under this benchmark scorer. The phases at the five selected snapshots are optimized with a batched Adam implementation. For each restart it uses 600 updates with a cosine learning-rate schedule beginning at 0.05 and ending near 0.0005. The first phase vector is zero and the other four are Gaussian with scale 0.5. The implementation keeps the best \emph{final} iterate of five restarts for each state and then the largest corrected score across the five screened snapshots. Ties are resolved in favor of the first occurrence. The fitted RZ angles and winning subrounds are retained in each control record of the present campaigns.

The phase random-number draws depend on the batch in which a candidate is scored, so a single call, a batch of five, and a larger neural batch can give different stochastic evaluations of the same candidate even with the same nominal seed. The shared code explicitly controls Python, NumPy, CPU-Torch, and CUDA-Torch streams where applicable, and stores the effective candidate slot, phase seed, batch shape, and result fingerprint. For the three-seed GP/classical comparison, the seed for the phase optimizer is computed from the UTF-8 byte sequence of the colon-separated ASCII string \texttt{00\_05:v1:phase:N:K:s:b}. The fields \(N,K,s,b\) are written as decimal integers: \(s\) is the whole-family optimizer seed and \(b\) is the zero-based first candidate slot in the submitted batch. We apply SHA-256 to that string, interpret its first four digest bytes as an unsigned big-endian integer, and take the result modulo \(2^{31}-1\). The string and byte-order rule fully specify the scalar seed; reproducing individual phase trajectories additionally requires the stated batch shape and pseudorandom-number generator versions. The neural campaign uses a different phase-seed schedule.

\subsection{Computational environment}

The validated five-seed campaign used Python 3.11.15, PyTorch 2.7.1 with CUDA 12.8, NumPy 2.4.6, SciPy 1.17.1, BoTorch 0.17.2, GPyTorch 1.15.2, and an NVIDIA A10G with approximately 22 GiB available device memory. The reference CUDA driver in that campaign was 595.71.05. Results on other GPU architectures or library builds need not agree bit for bit; the finite-precision checks above indicate the size of the numerical errors.

The reported runtimes include more than GPU kernels. A neural search can spend substantial time fitting and updating models and optimizing proposals. A GP search must update a probabilistic model and optimize its acquisition function after observing each batch. Nelder--Mead calls the scorer with a batch size of \(q=1\) candidate, which may use the GPU less efficiently than batches of \(q=5\). For this reason we report evaluation counts and wall time separately; the wall times include model training.

\section{Optimization settings and experimental accounting}\label{app:search}

\subsection{Neural architecture and proposal generation}

The source buffer covers all 15 targets with \(5\leq N\leq10\) and \(2\leq K\leq\lfloor N/2\rfloor\), with 50 observations per target. It combines 25 Latin-hypercube designs and 25 analytical warm-start points. The small-system design resets seed 42 for each target and retains the resulting repeated angle coordinates across distinct systems. The full data rows include \(N,K,\gamin,\gamsh\), and the corrected fidelity.

The neural ensemble comprises three residual MLPs with hidden widths \(384,512,320\), residual block counts \(4,5,6\), and dropout probabilities \(0.05,0.10,0.12\), plus two MoE networks with four and six experts of widths 256 and 320. The MoE gate uses \(N/20\) and \(K/N\) while expert predictions use the complete four-feature vector. Models are trained separately and all five contribute to the final proposal score. The high-fidelity weighting of Eq.~\eqref{eq:nnloss} is fixed throughout the compared protocols.

Initial model fitting uses at most 400 epochs, batch size 256, an 85:15 training/validation split, learning rate \(10^{-3}\), and weight decay \(10^{-5}\). The active-refinement schedule of the five-seed campaign uses continuation learning rate \(3\times10^{-4}\), patience 40, and maximum epoch caps \(120,135,150,165\) after its successive four refinement rounds. The lower/upper-confidence proposal rules use the mean and dispersion of Eq.~\eqref{eq:ensemble}. An uncertainty scale estimated from part of the training buffer serves as a ranking heuristic and carries no coverage guarantee.

Candidate optimization begins from multiple angle-space initializations and takes gradient steps through a subset of the ensemble before rescoring proposals with all five models. The proposal search uses a clamp \([0.02,\pi-0.02]^2\), slightly wider than the GP domain of Eq.~\eqref{eq:gp-domain}, approximately 400 optimization starts, 800 gradient steps, and candidate-search learning rate 0.015. To discourage unproductive boundary proximity, the gradient acquisition uses a smooth penalty
\begin{equation}
P(\boldsymbol\gamma)=
\sum_{a\in\{\mathrm{in},\mathrm{sh}\}}w_a
\left[e^{-\gamma_a/m}+e^{-(\pi-\gamma_a)/m}\right],
\label{eq:nn-penalty}
\end{equation}
with \(m=0.2\) rad, \(w_{\mathrm{in}}=0\), and \(w_{\mathrm{sh}}=0.15\). The method also keeps proposals diverse across ensemble members and across two shuttle-angle groups; the grouping only allocates the search and assumes no reflection symmetry in a single angle. The search pools candidates from ensemble members and confidence-based proposals, removes exact duplicates within each selection stage where prescribed, and prefers controls that are not clustered too closely in the two-dimensional angle domain. A selected pair adds a fidelity observation only after it has been scored by the oracle.

In the five-seed campaign, the neural budget is adaptive across targets. Total new target calls per 65-target execution are \(2325,2045,2140,2261,2278\) for seeds \(2026,9,11,919,20961\), respectively. The corresponding GP method pays exactly \(3250=65\times50\) new calls each run. The 750 source observations are reused in each neural run without being regenerated. The runtimes in Table~\ref{tab:five-seeds} are measured sequential wall times of the worker processes.

\subsection{Gaussian-process initialization, covariance, and acquisition}

The three-seed GP and classical studies use optimizer seed integers \(211,503,1009\). Each GP starts from five target-dependent candidate angles supplied by the approximate population-only warm-start model and five scrambled-Sobol coordinates spread across the allowed angle square. Those ten candidates are fully evaluated under the exact collision-and-phase oracle. Subsequent proposals depend on the observed scores, while all candidate and phase random streams are initialized deterministically from the system labels, optimizer seed, and candidate-batch position. Inputs are normalized from physical radians to the unit square. The fitted GP uses outcome standardization and a minimum noise floor \(10^{-4}\); the noise term accounts for the stochastic phase fit of the oracle, although the quantum evolution itself is noiseless.

The kernel is Mat\'ern-\(5/2\) [Eq.~\eqref{eq:matern}], with two numerical length-scale priors \(\Gamma(3,6)\) and output-scale prior \(\Gamma(2,0.15)\) in the fitted GP. The softplus parameterization starts the length scales at about 0.693147, which is neither the mean nor the mode of the prior. The ordinary and periodic models share these numerical conventions, but the effective local metric differs. In addition to the periodic distance of Eq.~\eqref{eq:periodic-distance}, a separate 12-target scale-control kernel uses
\begin{equation}
d_C^2(\boldsymbol\gamma,\boldsymbol\gamma')
=\sum_a\frac{\sin^2(\gamma_a-\gamma'_a)}
 {w^2\ell_a^2},
\label{eq:scale-control}
\end{equation}
which matches the ordinary GP's local distance asymptotically. This kernel was tested only on a fixed 12-target subpanel, so its three arms are not full-family replicates.

Each GP acquisition step selects five candidates using batch logarithmic expected improvement with 128 Monte Carlo samples, ten acquisition restarts, and 256 raw acquisition points. When a GP fit or an acquisition optimization fails, a fallback proposal is used and the failure is recorded. The fallback candidates are scored by the oracle and counted like any other evaluation. The three-seed full-family GP study makes \(65\times3\times50=9750\) new candidate evaluations per GP arm.

\subsection{Fixed and classical proposals}

The Sobol continuation inherits the identical ten GP initial observations for a target/seed pair and spends 40 new calls at unused scrambled-Sobol coordinates. The grid control instead pays 50 new calls for one analytical candidate and a fixed \(7\times7\) grid, using a prospectively specified order. This different initialization should be kept in mind when the grid and GP curves are compared at small budgets.

Bounded Powell and Nelder--Mead use two starting controls selected greedily among the best ten initial observations, separated by at least 0.25 radians. Each basin receives a 20-new-call budget. For Nelder--Mead, the initial simplex is formed from the starting point and axis-aligned offsets of magnitude 0.15 radians, reversing an offset when a forward step would exceed the allowed angle box. SciPy options are \(\mathtt{maxfev}=20\), \(\mathtt{maxiter}=1000\), \(\mathtt{xatol}=\mathtt{fatol}=10^{-4}\), and \(\mathtt{adaptive}=\mathtt{False}\). Powell uses the same 20-call cap with \(\mathtt{xtol}=\mathtt{ftol}=10^{-4}\). If one local solver terminates before spending its allocation, the remaining calls are assigned to additional preselected starting points or unused Sobol coordinates. A control submitted more than once is charged each time it is simulated.

The two-basin pattern search uses eight \(q=5\) batches, four for each basin. Its characteristic half-width on step \(h=0,1,2,3\) is
\begin{equation}
\delta_h=\frac{\pi-0.1}{4\cdot2^h}.
\label{eq:pattern-step}
\end{equation}
Four batch points are axis offsets from the current basin center; the fifth is a prescribed diagonal point. The basin center is updated using the best already scored point within that basin. Clipped and repeated points are charged as evaluations. Because this search repeats coordinates often, it should not be taken as representative of well-tuned pattern searches.

\begin{figure*}[t]
\centering
\includegraphics[width=0.86\textwidth]{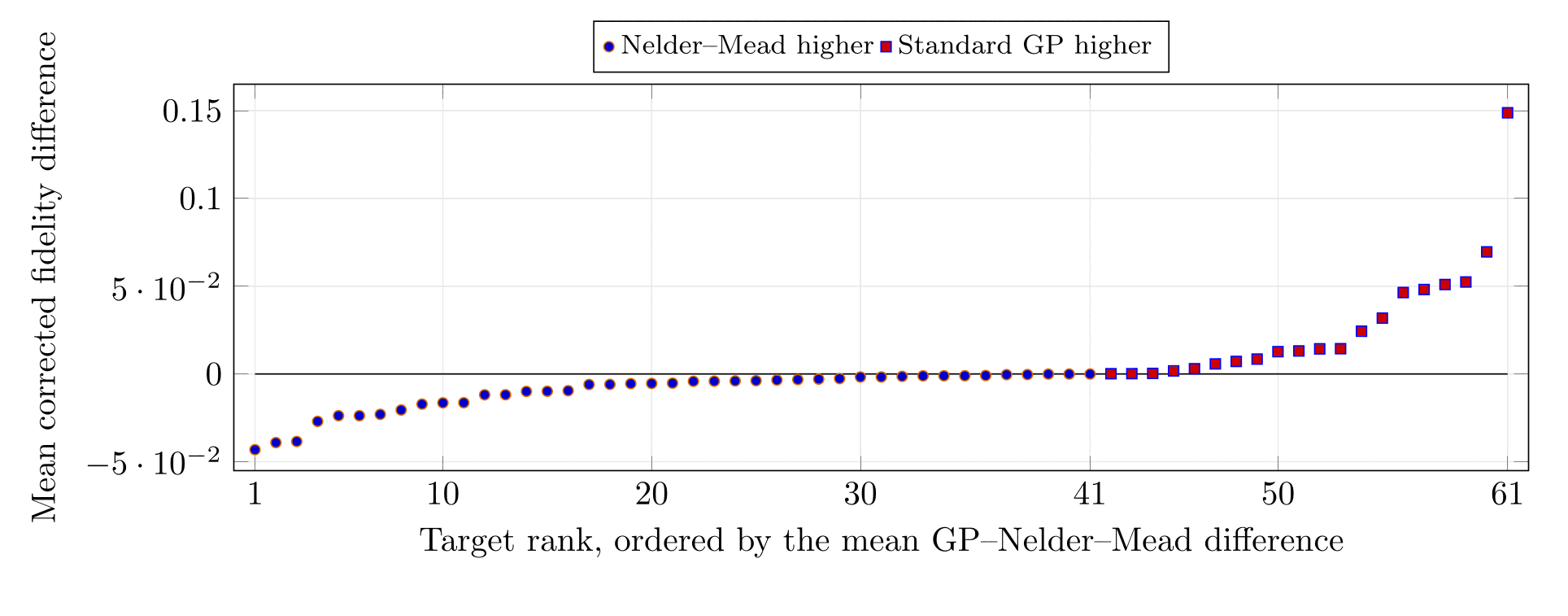}
\caption{Distribution of paired target-level mean fidelity differences between standard Gaussian-process Bayesian optimization and bounded Nelder--Mead at 50 logical angle evaluations. Each symbol represents one of the 61 targets with a composite-reference record; its plotted value is the average of the three corresponding optimizer-seed differences (seeds 211, 503, and 1009). Targets are sorted by this difference. Negative differences (circles, 41 targets) favor Nelder--Mead; positive differences (squares, 20 targets) favor GP. The mean of these differences is \(+0.002\) and the median \(-0.002\).}
\label{fig:gp-nm-targets}
\end{figure*}

\subsection{Low-label and source-only neural controls}

The source-ablation experiments compare a source-included arm, trained with the 750 small-system observations, with a source-excluded arm trained without them, holding the architecture, target train/validation membership, and newly evaluated target candidates fixed where required by the design. Both arms may pool observations across the 65 larger targets. Including the source rows increases the number of training mini-batches, so the two arms differ in training computation.

In the subsequent \(k=2\) and \(k=5\) low-label study, the initial target-angle lists are nested. The two-label list consists of the first analytical and first Sobol proposal; the five-label list extends it with one more analytical and two more Sobol candidates. The first active update brings both protocols to ten total observations, followed by increases to 20, 30, 40, and 50. The neural family seeds are \(2026,9,11\), with all target proposals prepared before the next full-family training update. The source-only experiment separately trains three fresh ensembles on the 750 source observations, selects five proposals per larger target without target labels, and compares them with five fixed angles produced by the same inexpensive population-only warm-start method, without using any measured large-target fidelities. Those source-only model weights are not used to initialize the low-label source-included and source-excluded arms.

The preliminary comparison of Bayesian variants was conducted on six selected targets, \((11,2),(11,5),(16,8),(18,8),(20,2),(20,10)\), with three independent seeds and six candidate-selection rules. The six rules were standard and periodic GP, two additional physically constrained periodic covariance models, GP without analytical initialization, and a Sobol continuation using the same analytical initialization. Each rule spent 50 oracle evaluations per target and seed, giving \(6\times3\times6\times50=5400\) main attempts. An additional 24 single-candidate calculations checked the collision-angle reflection symmetry under controlled phase initializations. These two parts account for the pilot's 5,424 newly executed calls. The pilot was used to select methods for the full-family study, and its results are not pooled with the later three-seed runs.

\subsection{Separation of campaigns, cost units, and statistical replicates}

\begin{table*}[t]
\caption{Overview of the numerical campaigns and their computational accounting. New angle evaluations denote full collision-angle candidates used in the optimization studies; the physical-dynamics and phase-certification studies instead count state trajectories and phase-fitting jobs. Historical observations reused across methods are not counted as newly evaluated candidates.}
\label{tab:campaigns}
\centering
\newcommand{\campStudy}[1]{\parbox[t]{.28\textwidth}{\raggedright\strut #1\strut}}
\newcommand{\campCalls}[1]{\parbox[t]{.18\textwidth}{\raggedright\strut #1\strut}}
\newcommand{\campScope}[1]{\parbox[t]{.46\textwidth}{\raggedright\strut #1\strut}}
\begin{tabular}{@{}lll@{}}
\toprule
\campStudy{Numerical study} & \campCalls{New angle evaluations} & \campScope{Scope}\\
\midrule
\campStudy{Neural surrogate versus standard GP}
& \campCalls{11,049 (neural) + 16,250 (GP)}
& \campScope{65 targets, five whole-family seeds; neural and GP calls recorded separately}\\
\campStudy{Bayesian pilot comparison}
& \campCalls{5,424}
& \campScope{Six selected targets, three pilot seeds, six controls; 5,400 main evaluations plus 24 diagnostic evaluations}\\
\campStudy{Standard versus periodic GP}
& \campCalls{21,300}
& \campScope{65 targets, three seeds, and a separate 12-target covariance-scale control}\\
\campStudy{Fixed sampling and source-data ablations}
& \campCalls{37,050}
& \campScope{65 targets, new fixed-design and source-ablation controls; previously evaluated GP observations reused}\\
\campStudy{Classical searches and low-label transfer}
& \campCalls{64,350}
& \campScope{65 targets, adaptive derivative-free controls and low-label/source-only tests; initial GP scores reused}\\
\campStudy{Time-resolved physical diagnostics}
& \campCalls{---}
& \campScope{59 unique 200-subround trajectories, 744 GPU standard snapshot fits, and eight stronger phase studies}\\
\campStudy{Independent phase-oracle certification}
& \campCalls{---}
& \campScope{Same 59 trajectories and 744 Adam fits as the row above, plus 921 coordinate-ascent fits and 16 L-BFGS cross-checks}\\
\bottomrule
\end{tabular}
\end{table*}

For a family with repeated optimizer seeds, we first average each method's results across seeds within a fixed target, and then average the target means equally. Paired differences are computed on the same target and seed only for arms designed as pairs, and the five-seed neural/GP and three-seed classical results are kept separate. Bootstrap intervals summarize a fixed, structured target family with three or five whole-family seeds and are used descriptively.

Source observations, reused initial candidates, new evaluations, phase-fitting restarts, and score-only applications are different computational resources. Neither the 750 source rows nor the extended multistart reference come with a full record of attempts. Together with the differences in horizon and phase fitting (Appendix~\ref{app:repro}), this rules out a quantitative speedup claim relative to Ref.~\cite{vu2026pra}.

\subsection{Five-seed neural and Gaussian-process results}

Table~\ref{tab:five-seeds} records the complete independent whole-family replicates underlying the five-seed comparison in Sec.~\ref{sec:results}. New call counts and elapsed times refer to all 65 target states in each execution, whereas the score averages use only the common 61-target reference intersection. Every neural run has access to the same 750 smaller-system observations.

\begin{table*}[t]
\caption{Paired five-seed neural-surrogate and standard GP-BO results. Each seed labels a complete 65-target execution. Fidelities and differences average the 61 common targets; new-call counts and times cover all 65 targets. The 750-observation source buffer, reused in every neural run, is not included in the new-call counts. Wall times include training, acquisition, oracle calls, and execution overhead.}
\label{tab:five-seeds}
\centering
\begin{tabular}{@{}rrrrrrr@{}}
\toprule
Seed & Neural mean & GP mean & Neural--GP & New neural calls & New GP calls & Time NN/GP (min)\\
\midrule
2026 & 0.766972 & 0.762343 & +0.004629 & 2325 & 3250 & 96.70 / 58.53\\
9 & 0.765839 & 0.761835 & +0.004004 & 2045 & 3250 & 91.74 / 57.86\\
11 & 0.765601 & 0.762639 & +0.002962 & 2140 & 3250 & 93.46 / 58.40\\
919 & 0.766096 & 0.763949 & +0.002146 & 2261 & 3250 & 96.51 / 58.18\\
20961 & 0.763000 & 0.764803 & \(-0.001803\) & 2278 & 3250 & 95.39 / 59.28\\
\midrule
Mean & 0.765501 & 0.763114 & +0.002388 & 2209.8 & 3250 & 94.76 / 58.45\\
\bottomrule
\end{tabular}
\end{table*}

\section{Phase-correction and time-resolved physical diagnostics}\label{app:phase}

\subsection{Local phase optimization as a nonconvex problem}

For a fixed state in the weight-\(K\) sector, define
\begin{equation}
S(\boldsymbol\phi)=
\sum_{|z|=K}a_z e^{i z\cdot\boldsymbol\phi},
\qquad F(\boldsymbol\phi)=\frac{|S(\boldsymbol\phi)|^2}{\Cnk}.
\label{eq:phaseS}
\end{equation}
The analytic gradient is
\begin{equation}
\frac{\partial F}{\partial\phi_i}
=-\frac{2}{\Cnk}\,
\mathrm{Im}\left[S^*\!
  \sum_{|z|=K}z_i a_z e^{iz\cdot\boldsymbol\phi}\right].
\label{eq:phasegradient}
\end{equation}
This objective has a global phase gauge: \(\boldsymbol\phi\mapsto\boldsymbol\phi+c\boldsymbol1\) multiplies \(S\) by \(e^{icK}\) and leaves \(F\) unchanged. A gauge-fixing convention may set the last phase to zero, but the phases still live on a \(2\pi\)-periodic torus, and a box boundary has no physical meaning.

For coordinate \(i\), separate the terms with \(z_i=0\) from those with \(z_i=1\):
\begin{align}
A_i&=\sum_{z_i=0}a_z
 e^{i\sum_{j\ne i}z_j\phi_j},\\
B_i&=\sum_{z_i=1}a_z
 e^{i\sum_{j\ne i}z_j\phi_j},\nonumber
\end{align}
so that \(S=A_i+e^{i\phi_i}B_i\). The squared modulus is
\begin{equation}
|S|^2=|A_i|^2+|B_i|^2+
2|A_iB_i|\cos(\phi_i+\arg B_i-\arg A_i).
\label{eq:coordinate-objective}
\end{equation}
When \(A_i\) and \(B_i\) are both nonzero, the exact maximizing update for that coordinate is
\begin{equation}
\phi_i\leftarrow \arg A_i-\arg B_i
\quad(\mathrm{mod}\ 2\pi).
\label{eq:coordinate-update}
\end{equation}
If either group vanishes, the coordinate is left unchanged. Each update can only increase the exact objective, but cycling through the coordinates gives no certificate of the global optimum. An update that lowers the recomputed score by more than the roundoff tolerance is rejected.

The phase certification uses a deterministic coordinate-ascent protocol with alternating forward and backward cyclic sweeps, at most 48 sweeps per start, and early stopping after two sweeps below a prescribed improvement threshold. The initial phase pool includes the zero vector, the original same-state final phase when available, the best direct application of the five already fitted top-five phase vectors, the nearest-time top-five phase vector, four uniform random vectors, and four independent Gaussian random vectors. Duplicate vectors can be omitted only under an exact gauge-equivalence rule. We retain the highest independently verified attained score as a lower bound on \(F_t^\star\).

The independent comparison uses a six-start L-BFGS-B phase fit with at most 150 iterations per start and an independently verified objective and gradient. Bounded phase representatives and local convergence criteria can yield different stationary values on the periodic torus. This cross-check provides a second numerical estimate; neither solver certifies \(F_t^\star\).

\subsection{Screened times versus a larger time shortlist}

The original oracle screens only the top five \(\Fabs\) subrounds. The certification was designed to investigate two possible sources of low fitted fidelity. First, the original Adam phase solver might converge to a poor stationary point even at a selected snapshot. Second, a state not among the five highest-amplitude-bound times might admit a better local phase correction than the chosen times. We test them at fixed angle pairs, without a new search over the collision angles. The analysis separates three effects:
\begin{enumerate}
\item \emph{Fitter-only:} compare the coordinate method with the original Adam on the same top-five snapshot set.
\item \emph{Snapshot screening:} with the coordinate method held fixed, compare the top-five set with the expanded set described below.
\item \emph{Combined effect:} compare the expanded coordinate method with the original top-five Adam result on the same newly simulated control.
\end{enumerate}

The expanded shortlist adds the fixed time anchors
\(\{0,1,5,10,25,50,100,200\}\) not already in the top five. It then applies the already fitted top-five phase vectors to every stored snapshot without new phase optimization. Among the remaining times it selects the three largest directly attained scores for additional coordinate fitting, using a deterministic earlier-time tie rule. Thus there are at most \(5+8+3=16\) distinct fitted snapshots per control.

The 59 unique controls in the prespecified physical panel yield 744 original Adam snapshot fits, 921 coordinate fits, and 16 independent six-start L-BFGS studies in the certification. Fifty-nine full 200-subround evolutions were performed anew because the previous study did not retain all dense state snapshots needed for arbitrary refitting. The duplicated control was reused. None of these jobs is a new collision-angle optimization.

The fitted snapshot comparisons yield 25 of 744 coordinate improvements above the numerical threshold of \(2\times10^{-6}\), including 21 gains larger than 0.01 and 14 larger than 0.05. Twelve snapshot/control pairs were selected for closer examination because an already-available phase vector from another subround attained a higher overlap than the original fitted vector on the same state. Each of these counterexamples needs only the direct application of a phase vector that had already been fitted. All twelve pointwise discrepancies are reproduced under independent scoring. In one especially clear case, \(\Dstate{14}{5}\) at subround 51, the original Adam phase result is \(0.001928896\), a previously obtained neighboring-time vector gives \(0.842799\), and coordinate ascent returns \(0.853954\) under the same state's \(\Fabs=0.864960\).

The original five-restart Adam trajectories at this state converge near a low-score stationary region. Their intermediate best scores differ from their final scores by at most numerical roundoff on the tested campaign; retaining earlier Adam iterates would not, by itself, repair these large failures. The zero-phase starting point takes a much better path under coordinate updates. Independent objective and gradient implementations reproduce both the low Adam value and the high attained coordinate value. We therefore attribute the discrepancy to the update path of the finite optimizer and the stationary regions it reaches; we found no indexing or fidelity-definition error.

Expanding the time shortlist under one corrected fitter produces 11 control-level gains above the numerical threshold, including five larger than 0.01. All five gains above 0.01 occur at the analytical or deliberately perturbed control angles in the fixed diagnostic panel, rather than the 36 previously optimized GP, neural, and Nelder--Mead controls. Across 36 selected GP/neural/Nelder--Mead champion controls, the mean combined gain is only \(6.92\times10^{-5}\), the maximum is \(0.001099\), and the final relative ordering is unchanged for all 12 target groups. These results support the stability of the selected controls in this panel. They do not show that every unsuccessful candidate was scored correctly or that all full-family rankings would survive a change of oracle.

\subsection{Independent scoring, remaining phase-fit uncertainty, and spin checks}

An independent complex128 NumPy implementation, an autograd implementation, finite-difference gradients, and explicit tensor-product \(R_Z\) tests agree under the stated qubit-bit convention. A final audit independently checks 35,511 numerical phase scores, all 9,964 frozen coordinate starts, 3,720 Adam restarts, 96 independent L-BFGS starts, and all 11,859 trajectory reductions. Its maximum independent stored-score discrepancy is \(4.02\times10^{-7}\), and recorded-state analytic/finite-difference gradient discrepancies remain below the prescribed tolerance.

The L-BFGS cross-check agrees with coordinate ascent within \(2\times10^{-6}\) on ten of the 16 selected states. On six it finds a lower, independently reproducible attained score, with differences between approximately 0.0022 and 0.0379. All such values remain above the direct phase vectors supplied as their starting lower bounds. We report these deficits as evidence that the attained phase-fit values depend on the heuristic solver. Neither solver proves global phase optimality.

For every computed state, an independently verified attained score \(L_t\) and the amplitude-only bound supply the exact enclosure
\begin{equation}
L_t\leq F_t^\star\leq\Fabs(t).
\end{equation}
At \(\Dstate{14}{5}\), subround 51, the width of this interval after coordinate ascent is about 0.011005; for other selected difficult snapshots it remains considerably larger. This width measures the remaining numerical uncertainty and should not be read as an irreducible frustration loss. The amplitude deficit \(1-\Fabs\) is exact for phase-only correction at that snapshot, whereas the maximum fidelity of the full protocol has not been bounded.

Independent small-\(N\) algebraic fixtures verify Eq.~\eqref{eq:spinweights} in 19 fixed-excitation cases through \(N=8\), with maximum spin-weight discrepancy on the order of \(10^{-15}\). Additional fixture states verify the invariance of Eq.~\eqref{eq:phase-plaquette}, including a deliberately incompatible uniform-amplitude \(\Dstate{4}{2}\) sector state with a phase of \(\pi\) on one occupied pair and zero phase on the other five pairs. These tests distinguish exact structural statements from numerical evidence at larger \(N\).

\subsection{Phase frustration and stopping-time observations}

The physical-dynamics study also sampled gauge-invariant plaquette phases from the time-resolved collision states. Across 59 unique fixed-control trajectories, the median, across trajectories, of the within-trajectory Spearman rank correlation between the sampled frustration statistic and the observed finite phase gap was approximately \(0.673\); 57 associations were positive, one was zero, and one negative. After a descriptive adjustment for amplitude and sample-support descriptors, 50 of 58 defined associations remained positive and eight were negative. The refitted phase values typically cover only 9--11 eligible time points per trajectory, and the plaquette sample does not exhaust all weight-\(K\) configuration relations. These associations describe the selected diagnostic states and should not be read as a causal law or a fidelity bound.

The winning subround also depends on the collision controls. For the 12 standard GP-BO controls, seven selected winners occur at \(t=K\), after the first complete cycle through the shuttles. The other five occur at later times. Neural and Nelder--Mead controls likewise exhibit both early and late winners. Because the ordered partial-SWAP gates do not commute in general, a contact count or a weighted scalar time alone does not characterize the evolution. A population-uniformity threshold can be crossed either before or after the best locally corrected fidelity.

A leave-one-system-size-out analysis tested whether elementary descriptors such as \(N\), \(K/N\), collision angles, and shuttle-contact counts predict the selected winner time. For the standard GP-BO controls, ridge regression gives a mean absolute error of 25.71 subrounds, compared with 27.03 for a training-median predictor; the target-resampled descriptive interval for the improvement crosses zero. Other tested angle and timing predictors also fail to establish a robust advantage over simple references. Among the tested descriptors and predictors, we thus find no evidence for a universal angle-scaling rule or a characteristic preparation time; models built on many-body observables may do better.

\section{Reproducibility and reference construction}\label{app:repro}

\subsection{Numerical origin of the comparison reference}

The collision-model study of Ref.~\cite{vu2026pra} treated Dicke targets containing up to 14 qubits and employed a dense multistart optimization of the two collision angles. We retain 31 optimized numerical records from the \(N=5\)--14 portion of that study, 16 of which concern the \(N=11\)--14 targets considered here. We supplement those results with 45 GPU-based numerical records obtained from an extended collision-angle search covering every target with \(N=15\)--20 and \(2\le K\le\lfloor N/2\rfloor\). Reference~\cite{vu2026pra} did not report these larger systems.

Table~\ref{tab:reference-values} lists all 61 reference values that overlap with our 65-target \(N=11\)--20 test family. Four smaller-system references are unavailable, as indicated by dashes at \((N,K)=(12,6),(13,6),(14,6),(14,7)\). Entries outside the defined excitation family are also marked by dashes. The arithmetic mean of the 61 available values is \(0.770529\), with every target weighted equally. This mean defines our composite reference; it was not reported in Ref.~\cite{vu2026pra}.

\begin{table*}[t]
\caption{Best-found RZ-corrected fidelities in the composite reference. Entries for \(N=11\)--14 come from the retained numerical results associated with the up-to-14-qubit search in Ref.~\cite{vu2026pra}. Entries for \(N=15\)--20 come from a GPU extension of that search carried out for this work. Dashes indicate unavailable reference values or excitation numbers outside the defined family; they are omitted from the 61-entry mean \(0.770529\). All entries are best-found values.}
\label{tab:reference-values}
\centering
\begingroup\scriptsize
\setlength{\tabcolsep}{4pt}
\begin{tabular}{@{}c*{9}{c}@{}}
\toprule
\(N\) & \(K=2\) & \(3\) & \(4\) & \(5\) & \(6\) & \(7\) & \(8\) & \(9\) & \(10\)\\
\midrule
11 & 0.868515 & 0.826315 & 0.912147 & 0.913520 & -- & -- & -- & -- & -- \\
12 & 0.861973 & 0.814285 & 0.887207 & 0.800176 & -- & -- & -- & -- & -- \\
13 & 0.853565 & 0.803376 & 0.813473 & 0.890520 & -- & -- & -- & -- & -- \\
14 & 0.849063 & 0.793436 & 0.756404 & 0.853559 & -- & -- & -- & -- & -- \\
15 & 0.841120 & 0.784330 & 0.743964 & 0.810395 & 0.845850 & 0.718411 & -- & -- & -- \\
16 & 0.837985 & 0.775951 & 0.732538 & 0.741777 & 0.839518 & 0.701670 & 0.706645 & -- & -- \\
17 & 0.830560 & 0.768210 & 0.722003 & 0.690238 & 0.806754 & 0.807852 & 0.687829 & -- & -- \\
18 & 0.828385 & 0.761031 & 0.712255 & 0.681043 & 0.734181 & 0.825452 & 0.662009 & 0.675958 & -- \\
19 & 0.821456 & 0.754348 & 0.703198 & 0.666104 & 0.674743 & 0.801296 & 0.775463 & 0.650226 & -- \\
20 & 0.819979 & 0.748109 & 0.694763 & 0.658035 & 0.633457 & 0.752582 & 0.804103 & 0.632979 & 0.640015 \\
\bottomrule
\end{tabular}
\endgroup
\end{table*}

The composite reference serves for orientation. The earlier paper used a maximum sweep count of \(R=200K\), while the new optimizers use \(T=200\) subrounds, and the angle optimization, phase fitting, and accounting of failed or repeated calls also differ. A small difference in best-found fidelity therefore cannot be assigned to the search algorithm alone, and the composite mean is not a performance limit of the collision circuit.

\subsection{Numerical software and retained observations}

All new simulations propagate the full, noiseless \(2^N\)-component state vector and evaluate the fixed-\(K\) Dicke-sector amplitudes. The target preparation, ordered two-body collision sequence, stopping-time selection, and final local-phase correction are defined in Sec.~\ref{sec:protocol} and Appendix~\ref{app:simulation}. The mathematical operational score is Eq.~\eqref{eq:operational-score}. The reported fidelity is thus fully defined in the text.

Across the five main optimization and comparison campaigns, we recorded 155,423 newly evaluated target--angle occurrences. Exact equality of the recorded tuples \((N,K,\gamin,\gamsh)\) reduces these to 102,297 distinct candidates, because individual searches can revisit controls and multiple experiments can query the same controls. Each recorded occurrence has the selected stopping subround and corresponding finite local-phase fit. Reused target observations, including the 750 smaller-system source labels and previously evaluated initialization controls, are distinguished from new oracle calls.

Exact reproduction of the stochastic optimizer traces requires the run journals, the phase-randomness and batching conventions described above, and the matching numerical software environment; their availability is summarized under Data availability.

\subsection{Numerical settings relevant to reproduction}

Table~\ref{tab:repro-settings} summarizes the core numerical settings used across the study. A reproduction should preserve the exact candidate order, phase RNG streams, seed-level train/validation splits, and batching conventions as well as the nominal model hyperparameters. These choices affect the finite oracle even though the collision dynamics are the same.

\begin{table*}[t]
\caption{Core numerical settings sufficient to identify the simulation and optimization protocols. The five-seed neural and three-seed classical comparisons are separate experiments.}
\label{tab:repro-settings}
\centering
\newcommand{\reproSetting}[1]{\parbox[t]{.28\textwidth}{\raggedright\strut #1\strut}}
\newcommand{\reproValue}[1]{\parbox[t]{.64\textwidth}{\raggedright\strut #1\strut}}
\begin{tabular}{@{}ll@{}}
\toprule
\reproSetting{Setting} & \reproValue{Value}\\
\midrule
\reproSetting{Target population} & \reproValue{\(N=11,\ldots,20\), \(K=2,\ldots,\lfloor N/2\rfloor\); 65 targets, 61 with composite reference values}\\
\reproSetting{Initial state and evolution} & \reproValue{\(\ket1^{\otimes K}\ket0^{\otimes(N-K)}\); 200 individual shuttle subrounds; complex64 state-vector simulation}\\
\reproSetting{Original local-phase correction} & \reproValue{Compute \(\Fabs(t)\) on all 201 snapshots; fit local \(R_Z\) on only the five highest-\(\Fabs\) snapshots using five Adam starts (one zero and four Gaussian), 600 steps, and initial learning rate 0.05 with cosine decay}\\
\reproSetting{Five-seed neural-versus-GP campaign} & \reproValue{\(2026,9,11,919,20961\); one complete 65-target execution per seed}\\
\reproSetting{Three-seed GP/classical campaign} & \reproValue{\(211,503,1009\); one complete 65-target execution per seed}\\
\reproSetting{GP acquisition} & \reproValue{Five analytical plus five scrambled-Sobol initial evaluations; eight batches of five, with 128 qLogEI Monte Carlo samples, ten acquisition restarts, and 256 raw starts}\\
\reproSetting{Neural initial training} & \reproValue{750 shared small-system observations, 85:15 train/validation split, AdamW with learning rate \(10^{-3}\), batch size 256, at most 400 initial epochs}\\
\reproSetting{Time-resolved physical diagnostics} & \reproValue{12 preselected target states and 60 fixed control labels (59 unique trajectories); all 201 snapshots reduced per trajectory}\\
\reproSetting{Local-phase certification} & \reproValue{59 trajectories, 744 standard Adam snapshot fits, 921 coordinate-ascent fits, and 16 independent L-BFGS studies}\\
\bottomrule
\end{tabular}
\end{table*}

For target-level learning comparisons, an optimizer seed is an entire repeated training and search run over the fixed target family. A single run may contain dozens of strongly related target scores. Quantitative statements about variation across seeds should therefore be based on the three or five independent whole-family seeds; target--seed pairs are not independent experiments.

\subsection{Replication}

To reproduce the numerical comparisons, the state-evolution gate order, candidate-angle initialization, optimizer seeds, phase random-number streams, model-training splits, and GPU batch grouping must all be fixed. Changing the phase-fitting method or the time-screening rule changes the operational objective. The phase-fitter study of Appendix~\ref{app:phase} is therefore reported as a separate validation, and the benchmark scores are left as originally measured.

\newpage

\bibliography{SD_refs}

\end{document}